\documentclass[amsmath,nobibnotes,aps,superscriptadress,amssymb,pra,aps,showpacs,superscriptaddress,twocolumn,longbibliography]{revtex4-2}
\pdfoutput=1
\usepackage[utf8]{inputenc}
\usepackage[english]{babel}
\usepackage[T1]{fontenc}

\usepackage{amsmath}
\usepackage{diagbox}
\usepackage{amssymb}
\usepackage{soul}
\usepackage{hyperref}
\usepackage{graphicx}
\usepackage{amsmath}
\usepackage{float}
\usepackage{enumitem}
\usepackage{tikz}
\usetikzlibrary{quantikz}
\usepackage{dsfont}
\usepackage{bbold}
\usepackage{cancel}

\usepackage{multirow}
\usepackage{subfigure}
\usepackage{mathrsfs}

\begin{document}

\author{Javier Gonzalez-Conde}
\email{\qquad javier.gonzalezc@ehu.eus}
\affiliation{Department of Physical Chemistry, University of the Basque Country UPV/EHU, Apartado 644, 48080 Bilbao, Spain}
\affiliation{EHU Quantum Center, University of the Basque Country UPV/EHU, Apartado 644, 48080 Bilbao, Spain}

\author{Zachary Morrell}
\affiliation{Advanced Network Science Initiative,
Los Alamos National Laboratory
Los Alamos, NM 87545, USA}

\author{Marc Vuffray}
\affiliation{Theoretical Division,
Los Alamos National Laboratory
Los Alamos, NM 87545, USA}

\author{Tameem Albash}
\affiliation{Center for Computing Research, Sandia National Laboratories, Albuquerque, New Mexico 87185, USA}

\author{Carleton Coffrin}
\email{\qquad cjc@lanl.gov }
\affiliation{Advanced Network Science Initiative,
Los Alamos National Laboratory
Los Alamos, NM 87545, USA}

\title{Cost of Emulating a Small Quantum Annealing Problem in the Circuit-Model}

\begin{abstract}
Demonstrations of quantum advantage for certain sampling problems have generated considerable excitement for quantum computing and have further spurred the development of circuit-model quantum computers, which represent quantum programs as a sequence of quantum gates acting on a finite number of qubits. Amongst this excitement, analog quantum computation has become less prominent, with the expectation that circuit-model quantum computers will eventually be sufficient for emulating analog quantum computation and thus rendering analog quantum computation obsolete.
In this work we explore the basic requirements for emulating a specific analog quantum computation in the circuit model: the preparation of a biased superposition of degenerate ground states of an Ising Hamiltonian using an adiabatic evolution. We show that the overhead of emulation is substantial  even for this simple problem. This supports using analog quantum computation for solving time-dependent Hamiltonian dynamics in the short term and midterm, assuming analog errors can be made low enough and coherence times long enough to solve problems of practical interest.
\end{abstract}

\maketitle
\newpage
\section{Introduction}

In recent years, circuit-model quantum computing has generated significant interest due to high profile experimental demonstrations of quantum advantage for specific computational tasks \cite{Arute2019,Wu_2021,Hoke2023}. In the circuit-model of quantum computing, quantum algorithms or programs are expressed as a sequence of unitaries acting on one or two qubits, and this paradigm is sufficient for reproducing all forms of quantum computation \cite{Deutsch1985,Barenco1995,Kitaev1997,Boykin2000,Shi2002,Aharonov2003}.
These recent demonstrations have largely overshadowed alternative models of quantum computation,
which we collectively call \textit{analog quantum computing}, whereby the computation is enacted by continuously evolving the quantum system according to a (possibly time-dependent) Hamiltonian as opposed to a discrete sequence of unitaries acting on a subset of qubits as in the circuit-model. While the different paradigms are computationally equivalent~ \cite{PhysRevLett.99.070502,Childs2009, aharonov2005adiabatic, equivalence}, the mapping between one to the other introduces overheads, which can make one approach more suitable than the other for performing specific computations.

It is widely believed that one of the first practical applications of quantum computers will be the simulation of the dynamics of quantum systems \cite{Lloyd_1996,Cirac2012,Nori2014,Preskill_2018, Tacchino_2019, Georgescu_2014,  Camps_2022, Clinton_2021, Efekan_2022, Berry_2007,  Childs2012, Barends_2016, Trotter_1959, Suzuki_1976,Layden2022, Pastori_2022, Dong_2021, Campbell_2019_rand,Low_2019,  Berry_2020, Watkins_2022,Chen_2021, Berry_2015, berry2007efficient, gonzalezconde2023mixed}.
The evolution of a quantum system according to a time-dependant Hamiltonian is a fundamental task that touches on a variety of fields spanning chemistry, condensed matter and high energy physics. Some important computations that take this form include Hamiltonian simulation \cite{Preskill_2018, Tacchino_2019, Georgescu_2014, Camps_2022, Clinton_2021, Efekan_2022, Berry_2007, Childs2012, Barends_2016, Trotter_1959, Suzuki_1976, Layden2022, Pastori_2022, Dong_2021, Campbell_2019_rand,Low_2019,  Berry_2020, Watkins_2022,Chen_2021, Chi2001, Farhi_2001}, adiabatic state preparation \cite{Sugisaki_2022,Coello_P_rez_2022, du2010nmr, wan2020fast, brierley2012adiabatic, kantian2010eta, farooq2015adiabatic, unanyan2001preparation, sorensen2010adiabatic} and adiabatic quantum computation \cite{qa_first_paper,qa_second_paper, Farhi_2000, boixo_model_paper, santoro2002theory, boixo2014evidence,johnson2011quantum,das2005quantum, hauke2020perspectives,morita2008mathematical,mbeng2019quantum,somma2012quantum, Albash_2018_2,pudenz2014error, Albash_2018, albash2015consistency,Itay_2015, vuffray2022programmable, nelson2021single,Nelson_2022,PhysRevApplied.19.034053, Kapit_2021}.
The problem is generally framed as solving the time-dependent Schr\"odinger equation,
\begin{equation}
i\hbar\frac{d}{dt}|\Psi (t) \rangle = \hat{H}(t) |\Psi (t) \rangle.
\label{eq:Sch_time}
\end{equation}
where $\hat{H}$ and $\ket{\Psi}$ are the time-varying Hamiltonian and the state, respectively.
In the circuit-model of quantum computing, simulation of this equation typically requires discretizing the time-evolution unitary, that is decomposing the dynamical process into a sequence of discrete time steps \cite{Camps_2022, Clinton_2021, Efekan_2022, Berry_2007, Childs2012, Barends_2016, Trotter_1959, Suzuki_1976, Layden2022, Pastori_2022, Dong_2021, Campbell_2019_rand, Low_2019, Childs_2018, Berry_2014, Berry_2015_2, Haah_2021, Kalev2021, _ahino_lu_2021,PhysRevX.11.011020,Martyn2023,Poulin2011}.
Many of these techniques focus on reaching an efficient asymptotic query complexity, which can hide large overheads in implementation that can affect real-world utility. 

In contrast to the circuit-model paradigm, the simulation of a time-varying Hamiltonian is the native operation for analog quantum computers, assuming the Hamiltonian to be simulated can be captured by the hardware-native Hamiltonian. This makes programming specifications simpler and can lead to much faster run-times \cite{somma2012quantum,Kapit_2021,R_nnow_2014}.

In order to demonstrate the challenges of simulating the dynamics of the time-dependent Schr\"odinger equation on a circuit-model system, this work considers a quantum simulation of quantum annealing~\cite{Farhi_2001,qa_second_paper,Farhi_2000} on four qubits. The task is to simulate the dynamics accurately enough to reproduce the state at the end of the evolution, which in the long-time limit is a non-trivial superposition of ground states of an Ising model \cite{Ising_1,Ising_2}. We focus on the second order Trotter-Suzuki formula and the Magnus expansion to implement our discretization. While other simulation algorithms exhibit better asymptotic complexity in terms of their general error bounds\cite{Low_2017,Low_2019,low2019hamiltoniansimulationinteractionpicture, Berry_2020, An2021timedependent}, product formulas have low overhead, making them suitable for hardware with limited resources, and have been shown to exhibit better than expected performance compared to what the general error bounds suggests~\cite{Average-Case,Blanes_1999,_ahino_lu_2021}, including for low-energy Hamiltonian simulation \cite{Changhao2021,Changhao2022,Kovalsky2023}. 

We consider two scenarios: (1) in order to assess the overhead associated with discretization only, we assume a closed system circuit-model device, where the only source of error is the approximation associated with discretization; (2) in order to assess whether faithful simulation can be performed in the presence of decoherence in spite of the discretization overhead, we assume physical open system models for the circuit-model device. For the latter, we provide a comparison against open system analog quantum annealing, where the overhead of discretization is absent. The lower overhead associated with the analog computation reduces the detrimental effects of noise, allowing for more accurate simulation over a longer range of simulation times.

Our results indicate significant run-time overheads for the discretization approach and a lack of resilience to open system effects even for this simple problem, indicating that circuit-model computers in the NISQ era \cite{Preskill_2018} will face significant challenges to simulating time-dependent Hamiltonian evolution when compared to an analog quantum computer. These findings are corroborated via simulations of open-quantum systems and commercial quantum computing hardware from IBM. 
Furthermore, in the early fault-tolerance era, these simple problems may provide useful benchmarks to assess the performance of devices with a small number of logical qubits.

The paper is organized as follows.  In Sec.~\ref{sec:background}, we give details of the Magnus expansion and Trotter-Suzuki method used for the discretization of the continuous-time dynamics, the quantum annealing protocol that we wish to emulate, and our metrics for certifying the result. In Sec.~\ref{sec:aqc-sim}, we give our results for the number of Magnus and Trotter steps required for high accuracy simulations.  In Sec.~\ref{sec:open}, we show how the accuracy of the simulation is affected when open system effects are taken into account. In Sec.~\ref{sec:analog}, we provide a performance comparison with analog simulation.  We conclude with a discission in Sec.~\ref{sec:discussion}.

\section{Discretized Simulation of Quantum Annealing}
\label{sec:background}
To simplify the forthcoming derivations, we assume a time-normalized physical model of the Schr\"odinger equation (we set $\hbar = 1$ and report the energy scale of the Hamiltonian in units of rad/ns),
\begin{equation}
i\frac{d}{ds}|\Psi (s) \rangle = T \hat{H}^{(P)}(s) |\Psi (s) \rangle , 
\label{eq:Sch_t}
\end{equation}
with $\hat{H}^{(P)}$ being the time-dependent application Hamiltonian of interest, $T$ being the total simulation time and $s = t/T$, $s \in [0,1]$ being the normalized evolution time. 

\subsection{Discretization of the time-dependent Schr\"odinger equation} \label{sec:Discretization}

In this section we introduce the basic concepts that we use for discretizing time-dependent Hamiltonian dynamics for simulation on a circuit-model quantum computer. In the circuit-model, any unitary transformation can be decomposed into a sequence of quantum logic gates \cite{Barenco_1995, Nielsen_Chuang}. In this sense, discretized quantum simulation aims to generate a unitary $\hat{V}$ based on a sequence of hardware-native logic gates such that $\hat{V}$ approximates the target unitary $\hat{U}$ to some desired accuracy $\epsilon$, i.e. as measured by the operator norm $\lvert\lvert \hat{U} - \hat{V}\rvert\rvert_{\infty} \leq \epsilon$. For us, the goal will be to approximate the time-evolution unitary, which we express as:
\begin{equation}
\footnotesize
    \hat{U} =  \prod_{k=1}^{N_M} \hat{U}(s_k,s_{k-1}) = \prod_{k=1}^{N_M} \mathrm{Texp}\left( -i T \int_{s_{k-1}}^{s_k} \hat{H}^{(P)}(s) ds \right) ,
    \label{eq:Usolution}
\end{equation}
where $N_M$ is the number of Magnus steps, $s_{k} =~k \delta$,  $\delta = 1/N_M$, and $\mathrm{Texp}(\cdot)$ denotes the time-ordered exponential.

In this work the implementation of Eq.~\eqref{eq:Usolution} on a circuit-model quantum computer is implemented using a combination of two basic techniques: the Trotter product formula \cite{Barends_2016, Trotter_1959, Suzuki_1976, Layden2022, Pastori_2022, _ahino_lu_2021, Childs_2021} and the Magnus expansion \cite{Magnus_expansion, Blanes_2009, Blanes_1999}. The Magnus expansion method allows us to approximate the time-ordered unitary as an expansion of locally constant terms in time for small intervals $\delta$. The Lie-Trotter series allows us to decompose the non-commuting Hamiltonian terms into a product of quantum gates. Together these techniques provide a method to approximate $\hat{U}$ in terms of a decomposition of single- and two-qubit quantum gates, with the approximation error controlled by the number of Magnus and Trotter steps.

\begin{figure}[t!]
\centering
\includegraphics[width=1.0\columnwidth]{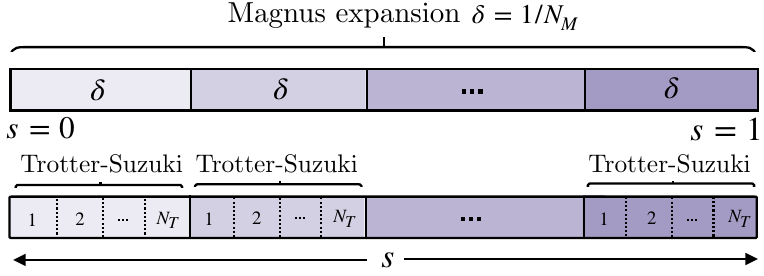}
\caption{An illustration of the discretization scheme used involving Magnus and Trotter-Suzuki expansions. The evolution from $s=0$ to $s=1$ is first discretized into $N_M$ segments, and each segment is approximated using a first-order Magnus expansion.  Each segment is then further approximated by $N_T$ Trotter-Suzuki steps.}
\label{fig:color_topo1}
\end{figure}

In the Magnus expansion the time evolution operator is represented as
\begin{equation}
\hat{U}(s_0+\delta, s_0) = \exp\left(-i\sum_{k=1}^\infty T^k \hat{\Omega}_k(s_0+\delta,s_0)\right) ,
\label{eq: magnus}
\end{equation}
with the first two terms in the expansion given by
\begin{eqnarray*}
\hat{\Omega}_1(s,s_0) &=& \int_{s_0}^s \hat{H}(s_1)\,ds_1,\\
\hat{\Omega}_2(s, s_0) &=& \frac{1}{2} \int_{s_0}^s ds_1 \int_{s_0}^{s_1} ds_2 \, [\hat{H}(s_1), \hat{H}(s_2)],
\end{eqnarray*}
with $[\hat{A}, \hat{B}] = \hat{A}\hat{B} - \hat{B}\hat{A}$ being the commutator. Truncating the expansion gives an approximation to $\hat{U}(s_0+\delta, s_0)$, with the quality of this approximation depending on the total simulation time $T$ and the time increment $\delta$. In order to keep the number of gates required low, we only consider the first term of the Magnus expansion, $\hat{\Omega}_1$, such that our approximation of the unitary $\hat{U}(s_0+\delta, s_0)$ is given by:
\begin{equation}
\hat{V}(s_0+\delta,s_0)= \exp \left(-i T \hat{\Omega}_1(s_0+\delta, s_0) \right) .
\end{equation}
For our Hamiltonian of interest, $\hat{\Omega}_1(s_0+\delta, s_0)$ is comprised of non-commuting Pauli operators, so we use the second order Trotter-Suzuki decomposition \cite{Suzuki_1976, Hatano_2005,berry2007efficient, zhuk2023trotter} to express $\hat{V}(s_0+\delta,s_0)$ as a product of gates acting on one and two qubits. For example, for $\hat{\Omega}_1(s_0+\delta,s_0) = \hat{A} + \hat{B}$ with $[\hat{A},\hat{B}]\neq 0$, 
the second order approximation using $N_T$ steps is given by
\begin{eqnarray} \label{eqt:TrotterSuzuki}
   \hat{V}(s_0+\delta,s_0) &\approx& \hat{V}_{\mathrm{TS}}(s_0+\delta,s_0)  =\left(e^{-i T \hat{B}/(2N_T)} \right. \nonumber \\
    && \left. \times \ e^{-i T \hat{A}/N_T} e^{-i T \hat{B}/(2N_T)} \right)^{N_T}  .
\end{eqnarray}
\begin{figure}[t!]
    \centering
    \subfigure[]{\includegraphics[width=0.25\textwidth]{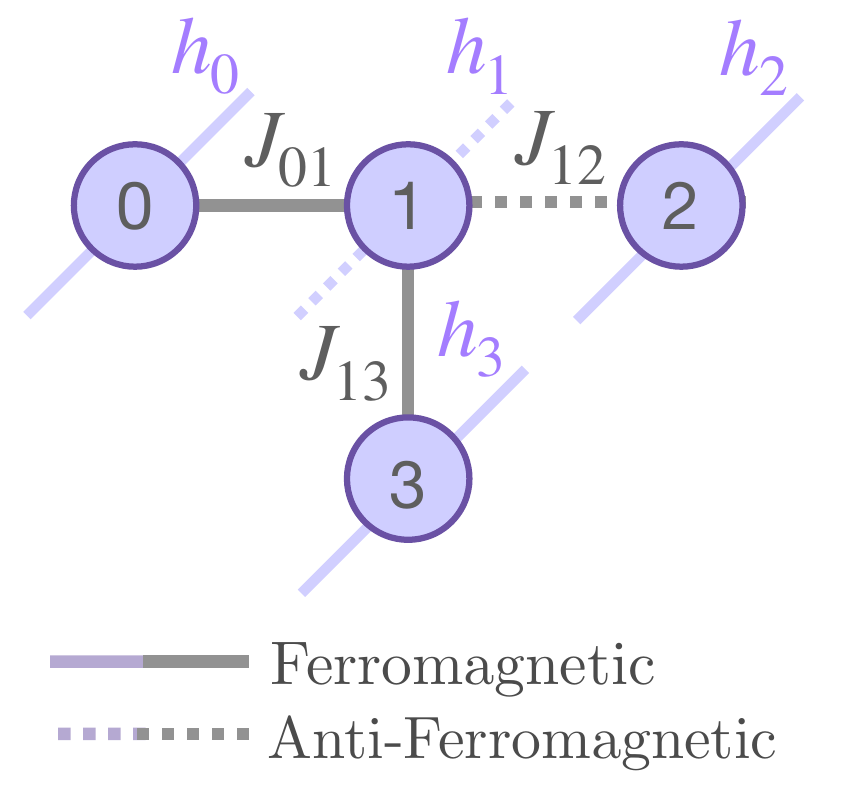}\label{fig:HT4-image}}
    \subfigure[]{\includegraphics[width=0.47\textwidth]{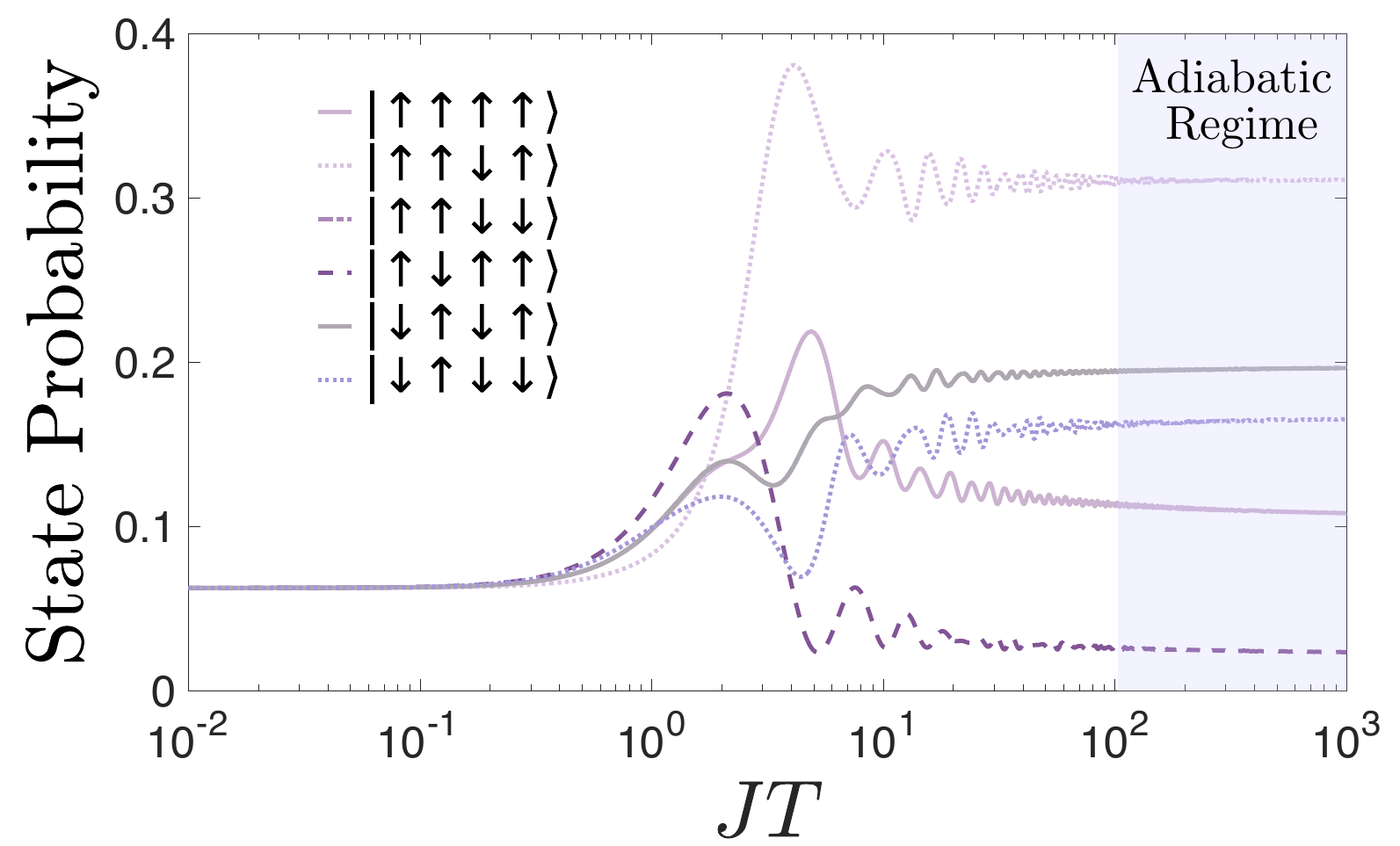}\label{fig:HT4_GSs_Populations}}
    \caption{(a) An illustration of the interaction graph of the four spin Ising model from $\hat{H}_{\mathrm{T}4}$ (Eq.~\eqref{eq:T_sig}) indicating ferromagnetic and antiferromagnetic interactions of strength 1. 
    (b) The populations of the Ising ground states at the end of the evolution as a function of annealing time $T$. }
\end{figure}
Increasing the order of the Trotter-Suzuki approximation improves the accuracy of the approximation for a fixed $N_T$ but at an increased overhead in the number of gates per Trotter step.
However, there is minimal extra overhead in using the second order formula as opposed to the first order formula since neighboring $e^{-i T \hat{B}/(2N_T)}$ gates can be combined into a single $e^{-i T \hat{B}/N_T}$ gate.
In our simulations, we employ the second order method to balance this trade-off.

In summary, we utilize a first order Magnus expansion and a second order Trotter-Suzuki decomposition to approximate the time-evolution unitary, Eq.~\eqref{eq:Usolution}, as a sequence of one and two qubit gates. The time-evolution unitary is discretized into $N_M$ steps, with each step approximated using the first-order Magnus expansion.  Each time step is then decomposed into $N_T$ second order Trotter-Suzuki steps. This procedure is summarized in Fig.~\ref{fig:color_topo1}. The individual single and two-qubit terms in each Trotter-Suzuki step can then be decomposed into the hardware-native logic gates of the circuit-model hardware.
The total runtime of the simulation is then given by the number of Trotter steps and the number of Magnus steps $N_M N_T$ and the quantum logic gates executed in each step.

\subsection{Quantum Annealing of Quantum Signature Ising Models}
\label{sec:qa-for-ising}
We focus on the requirements for simulating the time-varying Schr\"odinger equation in the context of  quantum annealing. The simulation task proceeds as follows. The initial state is given by an easily prepared ground state of a Hamiltonian $\hat{H}_\text{0}$. The state is evolved according to the parametrized Hamiltonian,
\begin{equation}
\hat{H}^{(P)}(s)=A(s)\ \hat{H}_0 +\ B(s)\ \hat{H}_\text{Target} ,
\label{eq:qa_ham}
\end{equation}
subject to the dynamical equation, Eq.~\eqref{eq:Sch_t}, for an amount of time $T$, which controls the speed of the evolution. 
The two interpolation functions $A(s)$ and $B(s)$ control the behavior of the dynamical system and are chosen such that $A(0) \gg B(0)$ and $A(1) \ll B(1)$. In this way, the interpolation starts with a Hamiltonian dominated by $\hat{H}_0$ and gradually  transitions to a Hamiltonian dominated by $\hat{H}_\text{Target}$. If the evolution satisfies the adiabatic condition \cite{adiabatic_cond} and in the absence of energy level crossings, the state at time $T$ will have high overlap with the ground state of the target Hamiltonian $\hat{H}_\text{Target}$.

One of the simplest and most well studied realizations of quantum annealing uses the transverse-field Ising model \cite{qa_first_paper,qa_second_paper, Stinchcombe1973IsingMI} to identify low-energy and ground states of classical Ising models, a problem which is NP-Hard in the most general case \cite{Barahona1982}. The Hamiltonian is specified using a graph with nodes $\mathcal{N}$ and edges $\mathcal{E}$ such that:
\begin{subequations}\small
\begin{align}
 \hat{H}_{0} &= -\sum_{i \in \mathcal{N}} \hat {\sigma}_i^x, \\
 \hat{H}_\text{Target} &= \hat{H}_\text{Ising}(J,h) = \sum_{(i, j) \in \mathcal{E}} J_{ij} \hat{\sigma}^{z}_i \hat{\sigma}^{z}_j + \sum_{i \in \mathcal{N}}  h_{i} \hat{\sigma}^{z}_i ,
\end{align}
\end{subequations}
where $\hat{\sigma}_i^\alpha$ denotes the $\alpha \in \{x,y,z\}$ Pauli operator acting on the $i$-th qubit, and the classical Ising model is defined by couplings and local fields $\left\{ J_{ij},h_i \right\}$. The initial Hamiltonian's ground state is the uniform superposition state, which is easy to prepare in practice. For simplicity, we consider a linear annealing schedule $A(s) = 1-s, B(s) = s$, although designing more complex schedules is a rich area of ongoing research with algorithmic advantages \cite{Grant_2021,Cote2023}.

Using the setup from Sec.~\ref{sec:Discretization}, each Magnus step of the Hamiltonian dynamics is approximated as,
\begin{eqnarray}
\footnotesize
\hat{V}(s_0+\delta,s_0)= \nonumber \\
&& \hspace{-2.8cm}  e^{-iT \left(\int_{s_0}^{s_0+\delta}A(s)ds \  \hat{H}_{0}+ \int_{s_0}^{s_0+\delta}B(s)ds\ \hat{H}_\text{Ising}(J,h)\right)}.
\label{eq:magnus_ising}
\end{eqnarray}
The Trotter-Suzuki expansion is then given by,

\begin{eqnarray}
\footnotesize
\hat{V}_{\mathrm{TS}}(s_0+\delta,s_0) &=&
\left(e^{-i \frac{T}{2 N_T} \left(\int_{s_0}^{s_0+\delta}A(s)ds \right)  \hat{H}_{0}} \right. \nonumber \\
&& \hspace{-1.5cm} \left. e^{-i \frac{T}{N_T}\left( \int_{s_0}^{s_0+\delta}B(s)ds \right)  \hat{H}_\text{Ising}(J,h)}\right. \nonumber \\
&& \hspace{-1.5cm} \left.e^{-i \frac{T}{2 N_T} \left(\int_{s_0}^{s_0+\delta}A(s)ds \right)   \hat{H}_{0}}\right)^{N_T}.
    \label{eq:trotter_ising}
\end{eqnarray}

An interesting class of Ising models that is used to benchmark quantum annealers are the so-called quantum signature Hamiltonians \cite{boixo_model_paper, Matsuda_2009}. These models have degenerate ground states with the distinctive feature that the adiabatic evolution of Eq.~\eqref{eq:qa_ham} does not prepare a uniform superposition of the classical ground states. Instead, these models feature a suppression and amplification of ground state probabilities, which can even vanish entirely in some cases \cite{boixo_model_paper}. For a given model, the probability distribution over the ground states can be predicted using degenerate perturbation theory.
While accurately reproducing the probability distribution of the ground states is a necessary but not a sufficient proof of the validity of the Hamiltonian simulation \cite{svmc, svmc_ibm} (see discussion in Appendix \ref{apx:SVMC}),
we focus on these benchmark problems because they provide a strong signal of the accuracy of the quantum simulation process even when restricting to the probability of computational basis measurement outcomes, which is a desirable property for benchmarking quantum computer hardware.

We use the following quantum signature Hamiltonian [depicted in Fig.~\ref{fig:HT4-image}] for our analysis,
\begin{eqnarray}
\hat{H}_{\mathrm{T}4}(s)&=&-(1-s)\left( \hat{\sigma}_0^x + \hat{\sigma}_1^x + \hat{\sigma}_2^x + \hat{\sigma}_3^x \right)\nonumber \\
&& +s\left(-\hat{\sigma}_0^z+ \hat{\sigma}_1^z  -\hat{\sigma}_2^z - \hat{\sigma}_3^z \right. \nonumber \\
&& \left. -\hat{\sigma}_0^z \hat{\sigma}_1^z  + \hat{\sigma}_1^z \hat{\sigma}_2^z - \hat{\sigma}_1^z \hat{\sigma}_3^z\right),
\label{eq:T_sig}
\end{eqnarray}
which we denote as the T$4$ Ising model. Here we assume that the Hamiltonian's energy scale is set to $J = 1 \mathrm{rad}/ns$. The Ising Hamiltonian of the above equation has six ground states [shown in Fig.~\ref{fig:HT4_GSs_Populations}], and we choose this model because it features a distinctive signature of Ising model ground state probabilities and can be implemented natively on a variety of quantum computing hardware.  We show the behavior of the ground state probabilities as a function of the total dimensionless evolution time $J T$ in Fig.~\ref{fig:HT4_GSs_Populations}. For very small evolution times (i.e., $\leq 10^{-1}$) the system remains close to the uniform super position state, where all computational basis states remain equally likely to be observed, so all ground states are equally populated. For medium evolution times (i.e., $10^{-1}$ to $10^{2}$) oscillations in ground state probabilities occur due to the effects of non-adiabatic evolution. For long evolution times (i.e., $\geq 10^{2}$) the system approaches the adiabatic limit and the ground state probabilities converge to a non-uniform probability signature due to the linear ${\hat{\sigma}}^x$ driver used in this demonstration.
\begin{figure}[t!]
\centering
\includegraphics[width=1\columnwidth]{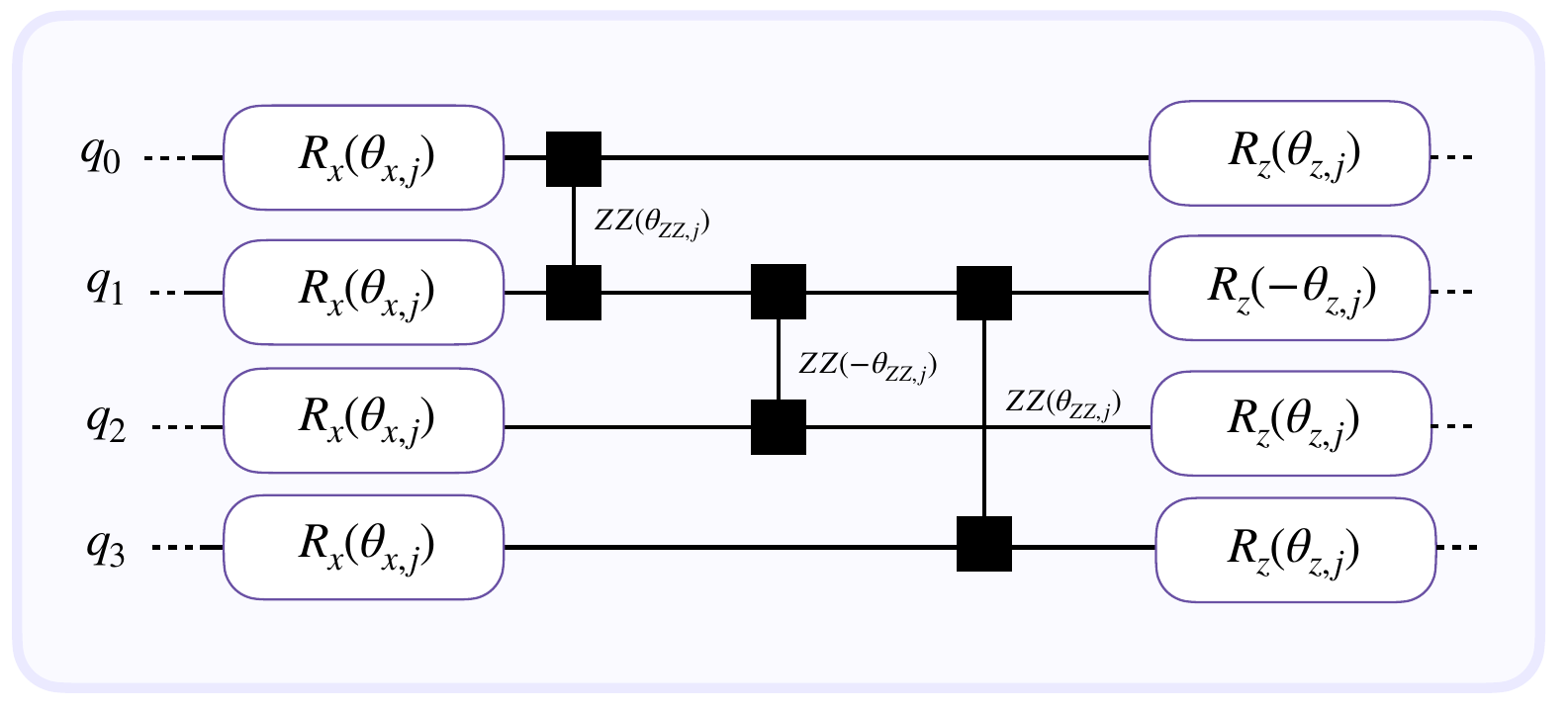}
\caption{Circuit implementation of a single Trotter step of the $j$-th Magnus step with a second order Trotter-Suzuki discretization of the Hamiltonian $\hat{H}_{\mathrm{T}4}$. $R_\alpha(\theta) = \exp(-i \theta \hat{\sigma}^{\alpha}/2)$ denotes a single-qubit rotation about the $\alpha$-axis by an angle $\theta$. $ZZ(\theta)$ denotes the two-qubit unitary $\exp(-i \theta \hat{\sigma}^z_1 \hat{\sigma}^z_2/2)$, which can be implemented with two CNOT gates and a rotation about the $z$ axis.  
Since single qubit rotations from adjacent Trotter steps can be combined, we only depict one layer of the $R_x$ gates.}
\label{fig:color_topo}
\end{figure}
\subsection{Gate set} \label{sec:GateSet}
With this fixed choice of Hamiltonian, we can complete our discretization of the quantum simulation. For simplicity, we assume access to arbitrary single-qubit rotation gates and two-qubit controlled-NOT (CNOT) gates. While any unitary can be exactly decomposed into these gates, this is \textit{not} a finite gate-set, so we do not refer to this as a digital quantum computation \cite{Deutsch2020}. The ability to do arbitrary single qubit rotations as opposed to approximating them from a finite universal gate set means that we will be \textit{underestimating} the overheads of our discretized quantum simulation. Therefore, our runtime estimates should be taken as a lower bound on the true cost for implementing the simulation using a finite universal gate set. 
Furthermore, we will assume arbitrary connectivity between the qubits, so there will be no additional overhead arising from implementing SWAP gates.  The hardware platforms we test on will also have the necessary connectivity to avoid implementing SWAP gates.

Under these assumptions, the discretization of the unitary operator $\hat{V}_{\mathrm{TS}}(s_0 + \delta,s_0)$ [Eq.~\eqref{eq:trotter_ising}] associated with $\hat{H}_{\mathrm{T}4}(s)$ is given in Fig.~\ref{fig:color_topo}. The simplicity of the circuit highlights how straightforward this particular simulation task is in terms of minimal compilation overhead.

\subsection{Simulation Validation}
For our proposed quantum simulation task, we measure the accuracy of the simulation in multiple ways. The most stringent metric we use is the infidelity between the computed state $\ket{\Phi}$ and desired state $\ket{\Psi(1)}$,
\begin{equation}
1-F = 1-| \braket{\Psi(1)}{\Phi} |^2.
\end{equation}
\begin{figure}[tb!]
    \centering
    \subfigure[\ $J T = 10^2$]{\includegraphics[width=0.475\textwidth]{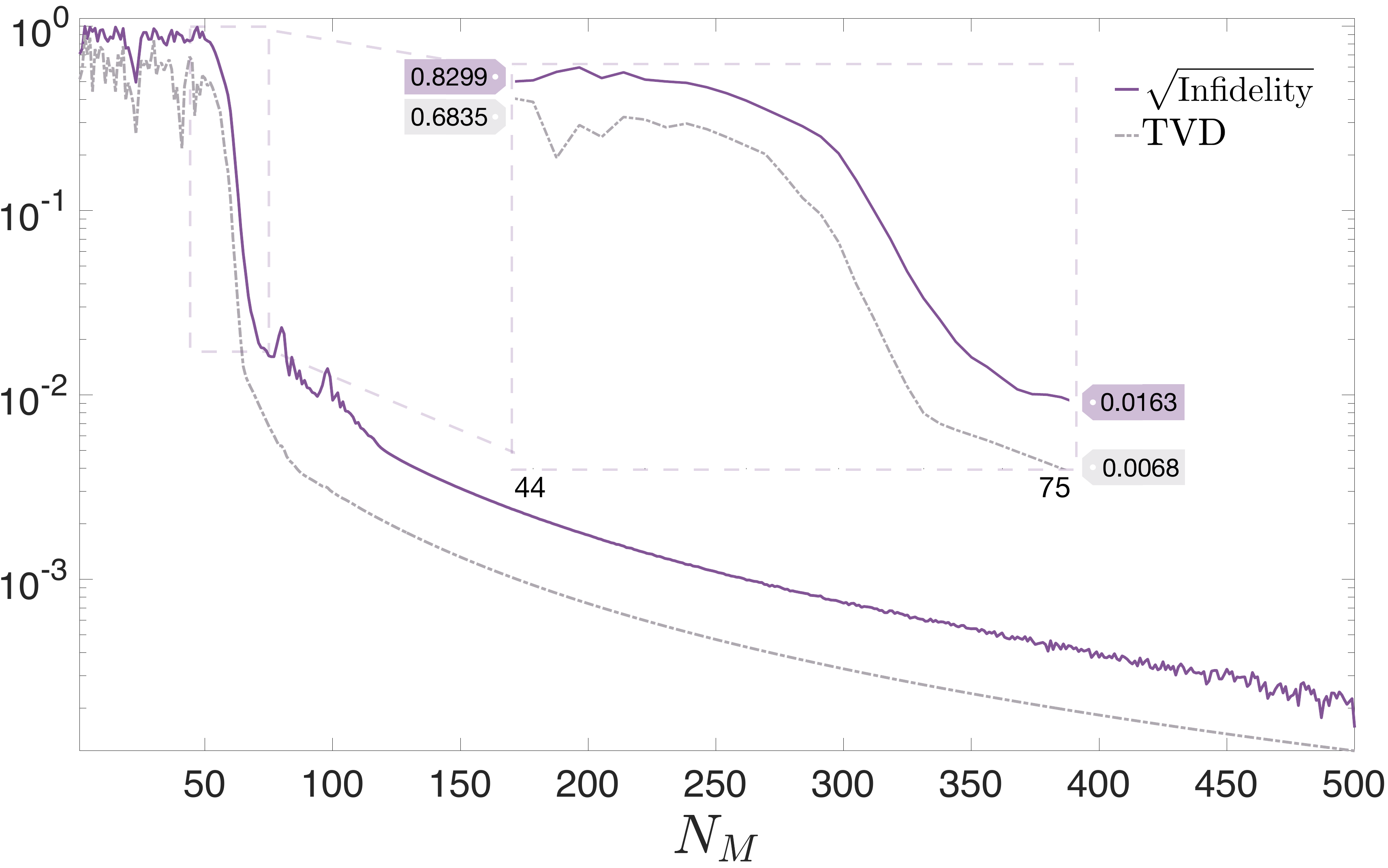}\label{fig:100ns} }
    \subfigure[\ $J T = 10^3$]{\includegraphics[width=0.465\textwidth]{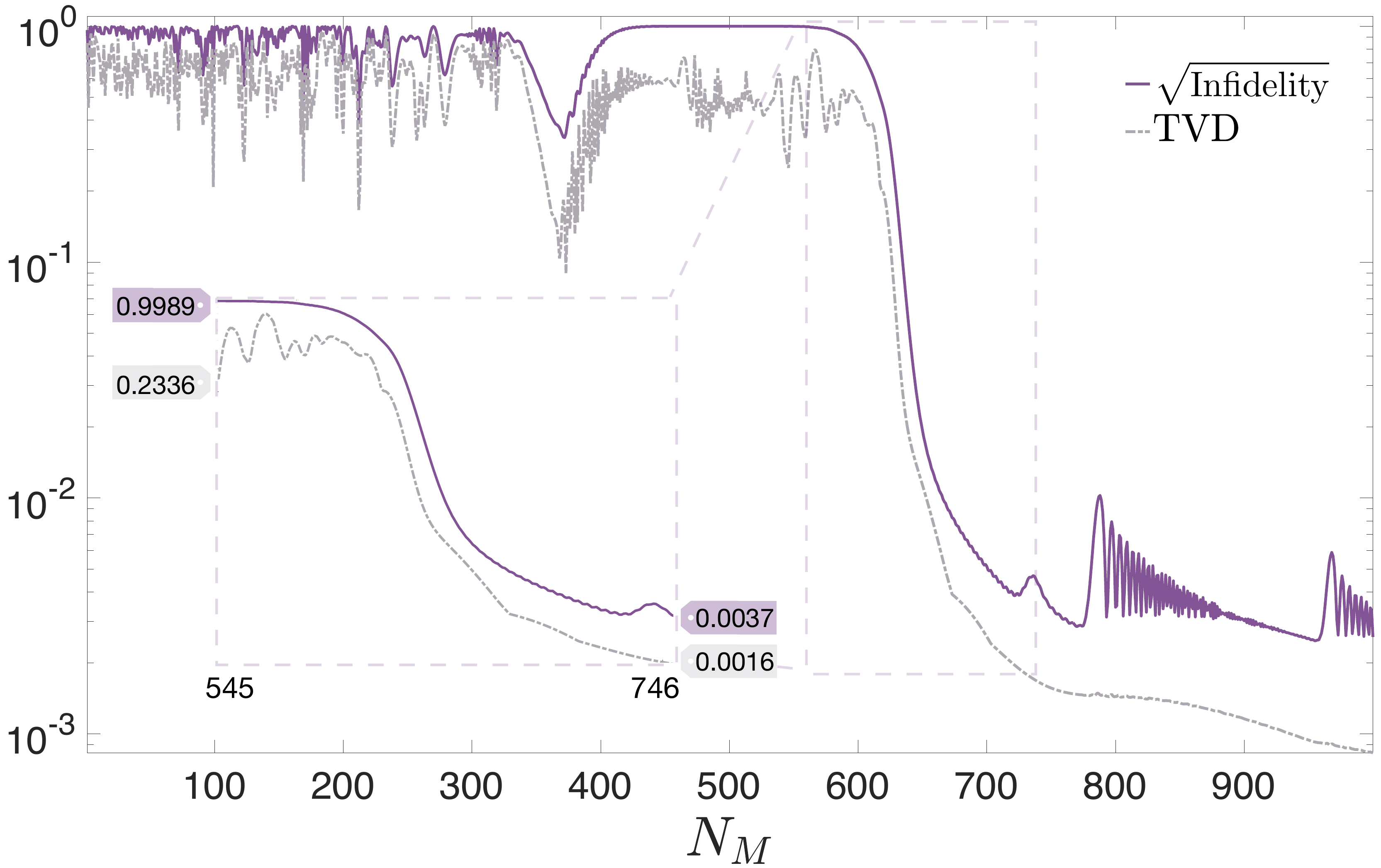}\label{fig:1000ns}}
\caption{Comparing the state infidelity and TVD between the exact and discretized dynamics with two Trotter steps for two different simulation times, (a) $J T = 10^2$ and (b) $J T = 10^3$, as a function of the total number of Magnus steps $N_M $. }
\label{fig:TVD_Infidelity}
\end{figure}
For our numerical simulations, the fidelity $F$ is calculated using the {exact} solution for the continuous-time dynamics of the model obtained using the Julia software package QuantumAnnealing \cite{julia_QA}.  The package uses a 4-th order Magnus expansion and increases the discretization until two conditions are reached: the $L_{\infty}$ norm between the density matrices at successive step sizes is less than $10^{-4}$ and the $L_{1}$ norm between the density matrices at successive step sizes is less than $10^{-6}$.

On physical quantum hardware, calculating the fidelity is far more challenging, for example requiring quantum state tomography \cite{dariano2003quantum}. Furthermore, some analog computing platforms may not have the flexibility (or give the user the ability) to perform measurements in different bases.
Therefore, to facilitate our study of the performance of quantum hardware across different platforms, we choose to use the total variation distance (TVD) between probability distributions induced by measuring the simulated states in the computational basis, defined as:
\begin{equation}
    \text{TVD}\;(p,q) = \frac{1}{2} \sum_i |p_i - q_i| , \\ 
\end{equation}
where $p_i$ and $q_i$ denote the $i$th computational basis populations for the simulated and ideal state respectively. The prefactor of $\frac{1}{2}$ normalizes the TVD to the range $[0, 1]$, which is why this metric is often presented as a percentage between $0\%$ and $100\%$. For the closed quantum system simulations, reproducing the correct population distribution of the Ising ground states is strongly correlated with the state infidelity (see Fig.~\ref{fig:TVD_Infidelity}), making the TVD an inexpensive proxy for the simulation accuracy. 
However, more care is needed in interpreting the TVD since it fails to capture coherence between Ising ground states. For example, a mixed state that reproduces the correct populations of the Ising ground states but does not reproduce the coherence between them would have the same TVD as a quantum state that also capture the coherences. We give an example of this in Appendix~\ref{apx:SVMC}. Nevertheless, a poor TVD can always be interpreted as a poor emulation of the true dynamics, and this is sufficient for our purposes.

For the purpose of this work, we treat any comparison under 1\% TVD to be {high-quality} and essentially the two distributions are  indistinguishable. We see from Fig.~\ref{fig:TVD_Infidelity} that how this TVD is reached depends strongly on the simulation time $JT$.  For large simulation times, the TVD is mostly flat until a sufficiently large number of discretization steps is reached, at which point it drops rapidly and continues to decrease with increasing number of steps. This large number of steps makes accurate simulations in the presence of noise prohibitive with current hardware, as we demonstrate in Sec.~\ref{sec:open}.

\section{Discretization Requirements for Simulating Quantum Annealing}
\label{sec:aqc-sim}
%
Using the quantum simulation task as well as the discretization procedure discussed in the previous section, our estimate of the discretization requirements is determined by the number of Magnus and Trotter steps required to produce a high-quality simulation. 

Assuming the number of Magnus steps is sufficient to achieve the desired accuracy, the general Trotter-Suzuki bound provides a worst-case upper-bound for the number of required steps for Eq.~\eqref{eqt:TrotterSuzuki} to achieve a specific accuracy \cite{Layden2022, Kivlichan_2020} 

\begin{table*}
\centering
\begin{tabular}{r||r|r|r|r|r|r}
 Simulated    & Magnus & Trotter & Total & Theoretical  &TVD & Fidelity  \\
  Time ($J T$)  & Steps & Steps & Steps & Bound \eqref{eq:order:trotter2} & & \\
\hline
\hline
0.01 & 1 & 1 & 1 & 1& 0.0001 & 0.9999 \\
\hline
0.10 & 1 & 1 & 1 & 1& 0.0053 & 0.9999 \\
\hline
1.00 & 5 & 1 & 5 & 5& 0.0075 &  0.9999 \\
\hline
10.00 & 17 & 1 & 17 & 56&0.0093 &  0.9996\\
\hline
100.00 & 70 & 2 & 140 & 741& 0.0095 & 0.9995 \\
\hline
1000.00 & 660 & 2 & 1320 & 7639& 0.0082 & 0.9989 \\
\hline
\end{tabular} 
\caption{The number of steps required to achieve a TVD accuracy of 1\%  on a circuit-model quantum computer as described in Sec.~\ref{sec:GateSet}. To calculate the theoretical bound we fix the number of Magnus steps to the values in the second column and use the Trotter-Suzuki bound in Eq.~\eqref{eq:order:trotter2} to calculate the number of Trotter steps. The theoretical bound on the total number of steps is then the product of the two quantities.} 
\label{tb:ht4-mag-trot}
\end{table*}
\begin{figure}[b!]
    \centering
    {\includegraphics[width=.47\textwidth]{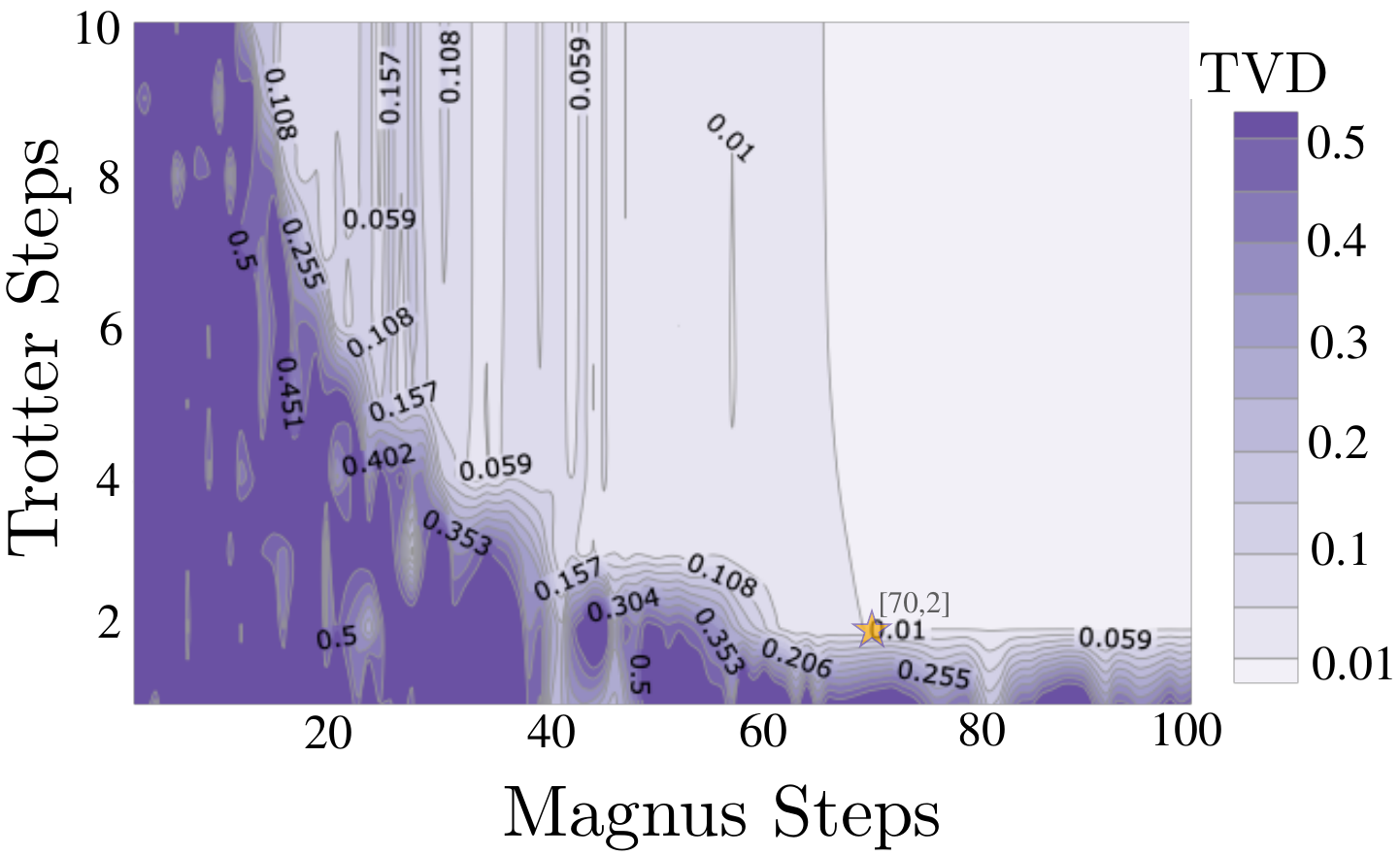} }
\caption{TVD for different choices of the number of Magnus and Trotter steps for a simulation time of $JT = 10^2$.  The ``star'' corresponds to a choice that achieves a TVD $<10^{-2}$ that minimizes $N_M N_T$ at $[N_M=70,N_T=2]$. It corresponds to the value reported in Table~\ref{tb:ht4-mag-trot}}.
\label{fig:TVD_Search}
\end{figure}
\begin{equation}
\footnotesize
    \mathcal{O}\left(\sqrt{\frac{(T/N_M)^3 (||[[\hat{A},\hat{B}],\hat{B}]||+0.5||[[\hat{A},\hat{B}],\hat{A}]||) }{12\epsilon}}\right).
    \label{eq:order:trotter2}
\end{equation}
Because this may be an overestimate of the number of steps needed, we instead minimize the number of total discretization steps $N_M N_T$ empirically to achieve a desired accuracy of 1\% in the TVD measure. As an example, we show in Fig.~\ref{fig:TVD_Search} the empirical TVD for different choices of Magnus and Trotter steps for $JT = 10^2$. We provide the figures for other values of $JT$ in Appendix~\ref{app:TVDsearch}. 

The identified minimal values for the total number of steps are presented in Table \ref{tb:ht4-mag-trot}.  Because the bound in Eq.~\eqref{eq:order:trotter2} does not provide the necessary constants for a direct comparison, in order to compare these values to the bound we assume the same number of Magnus steps and use the bound to provide an estimate for the number of Trotter steps needed to achieve our desired error.  Further details are provided in Appendix \ref{apx:error_bounds}. As shown in Table \ref{tb:ht4-mag-trot}, the theoretical estimate of Eq.~\eqref{eq:order:trotter2} significantly overestimates the total number of steps at large simulation times. Moreover, the total error is asymptotically driven solely by the total number of Magnus steps (see Appendix~\ref{apx:laTM}). This explains why the observed optimal choice of Trotter steps is so low.

\section{Benchmarking Quantum Annealing Simulations on NISQ Hardware} \label{sec:open}

Having established the discretization parameters needed to simulate quantum annealing accurately at different annealing times (see Table~\ref{tb:ht4-mag-trot}), we now consider the impacts of different forms of noise on these computations.

Thus far the task of simulating the Schr\"odinger equation with a quantum computer has been presented assuming that the quantum computer is a closed quantum system. However, current quantum computing devices are impacted by a wide variety of operational imperfections and noise from the surrounding environment \cite{vuffray2022programmable,PhysRevApplied.19.034053,Caldeira1981,Breuer2002,Flannigan_2022,9951204}.
This noise not only impacts the accuracy of the quantum gates implemented on the hardware but also the maximum duration of computations that can be executed. The precise computational failure modes that are induced by noise are platform and application dependent.

To provide insights into the impacts of noise, we compare three cases: numerical simulations of ideal closed-system quantum computations; numerical simulations of noisy open-system models; and demonstrations on current quantum computing hardware.

\subsection{Open System Models}
In order to numerically simulate the relevant open system dynamics, we leverage the noisy quantum simulation software from IBM's qiskit library called \textit{FakeMumbai} \cite{qiskit}, which strives to be a reasonable approximation of the dynamics of IBM's superconducting qubit quantum computers. By using the method  \textit{from\_backend()}, the open system model allows us to include the effects of several quantum channels: readout error, depolarizing error and thermal relaxation error. These models are constructed from the calibration data provided by the file \textit{props\_{mumbai}\_1.4.5.json} \cite{mumbai_qiskit_mock_local}. 

In our simulations, we consider two different open system models.  The first (noisy discretized 1) includes only the effect of decoherence in terms of phase damping error, characterized by the $T_2$ time. We have chosen the calibration values of \texttt{ibmq\_mumbai} version 1.4.5 \cite{mumbai_qiskit_mock_local}. The second (noisy discretized 2) includes the effect of all the noise channels described above in the hope to best replicate the behavior of the physical hardware. Details of the parameter choices for these models are given in Appendix \ref{apx:lima}. 

\subsection{Solution Quality}

\begin{figure}[tb!]
    \centering
    \subfigure[]{\includegraphics[width=0.485\textwidth]{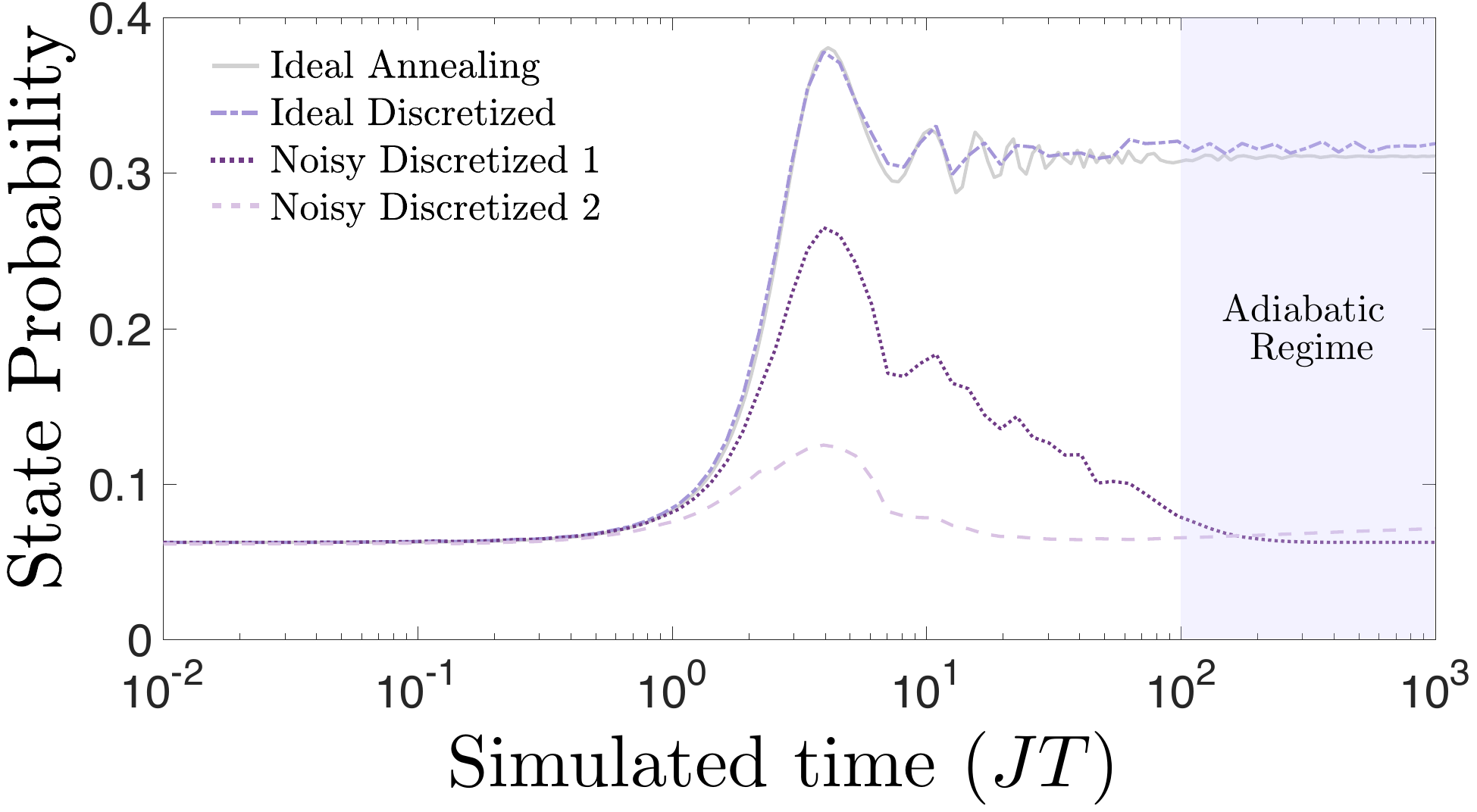} \label{fig:StateProbabilityOpen}}
    \subfigure[]{\includegraphics[width=0.489\textwidth]{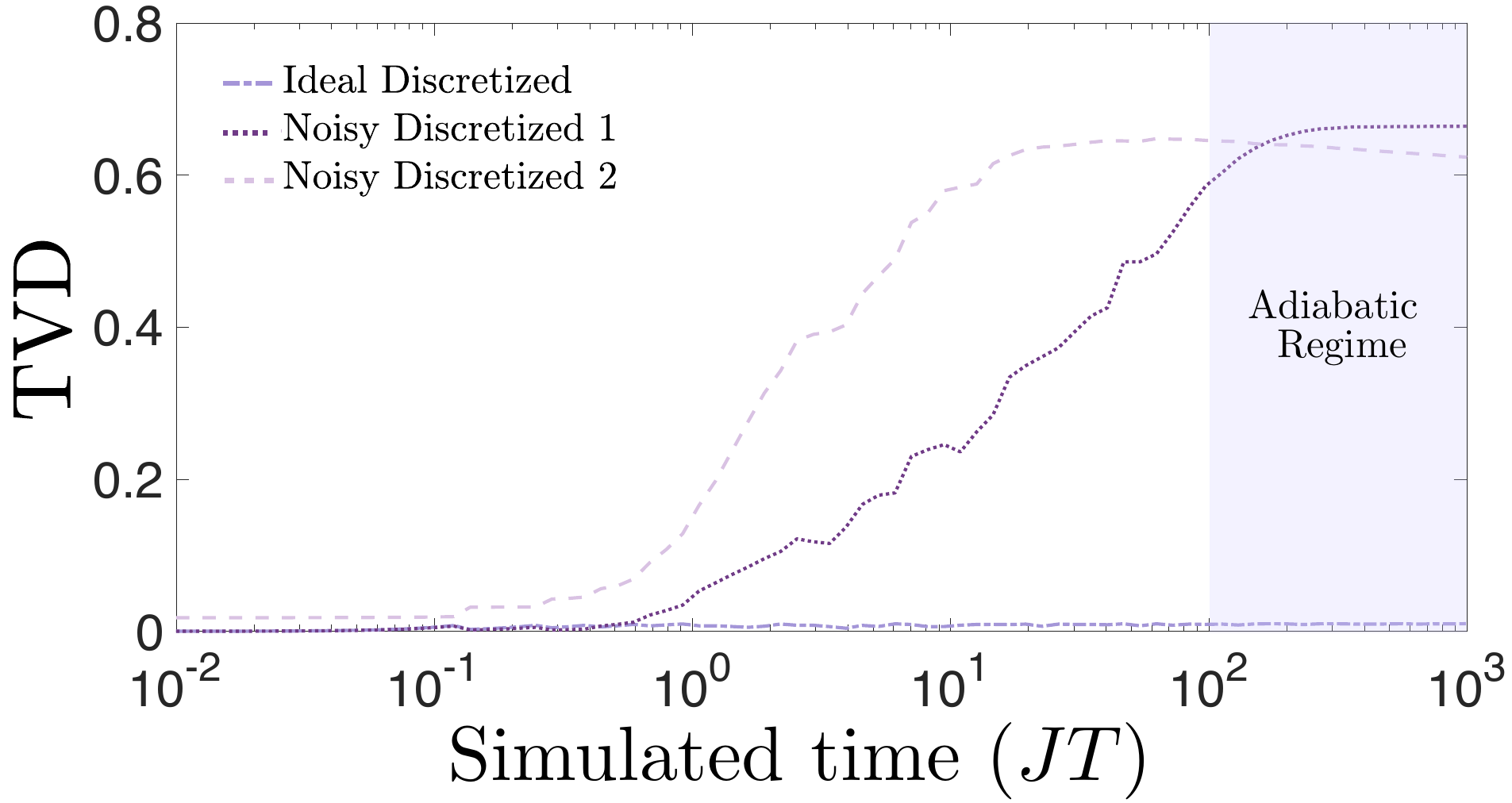} \label{fig:TVDOpen}}
    \caption{An analysis of the impact of open system effects on the simulation quality as a function of simulation time $J T$. Ideal Discretized corresponds to simulations with  $N_T = 2$ and the minimum number of Magnus steps $N_M$ needed to achieve a TVD $< 0.01$ in the ideal discrete case. Noisy Discretized 1 are open system simulations with only phase damping as described in the main text. Noisy Discretized 2 are open system simulations with readout error, depolarizing error and thermal relaxation error as described in the main text. (a) Probability of the ground state $\ket{\uparrow \uparrow \downarrow \uparrow }$ signature from $\hat{H}_{\mathrm{T}4}$, which should converge to a state probability of 1/3 in the long-time limit. For the probabilities of the other ground states see Fig. \ref{fig:7_appendix} in Appendix~\ref{app:additional}.
(b) TVD for the full state distribution produced and includes a uniform distribution as a point of reference. }\label{fig:Open}
\end{figure}

We show in Fig.~\ref{fig:Open} results comparing the open system simulations to the closed system simulations as a function of total evolution time.  As an illustrative example, we show in Fig.~\ref{fig:StateProbabilityOpen} how the probability of finding one of the ground states changes as a function of simulation time in the presence of open system effects.  For the decoherence parameters we have used, the simulation fails to track the true probability at approximately $J T = 1$.  This is reflected in the TVD in Fig.~\ref{fig:TVDOpen}, where we see that the error is already rising at this simulation time value. As we established in Sec.~\ref{sec:qa-for-ising}, this amount of evolution time is still very far from the adiabatic time scale, indicating that high quality simulation of long-time quantum annealing computations presents a challenge for current noisy circuit-model quantum computers.

In terms of the TVD, the difference between the two open system models also becomes more manifest at $J T > 1$. The TVD of the more comprehensive noise model (noisy discretized 2) rises more quickly, indicating that the other sources of error play a role too on top of dephasing errors. Nevertheless, dephasing errors appear to capture a significant fraction of the errors.

\begin{figure}[b!]
    \centering
    {\includegraphics[width=.485\textwidth]{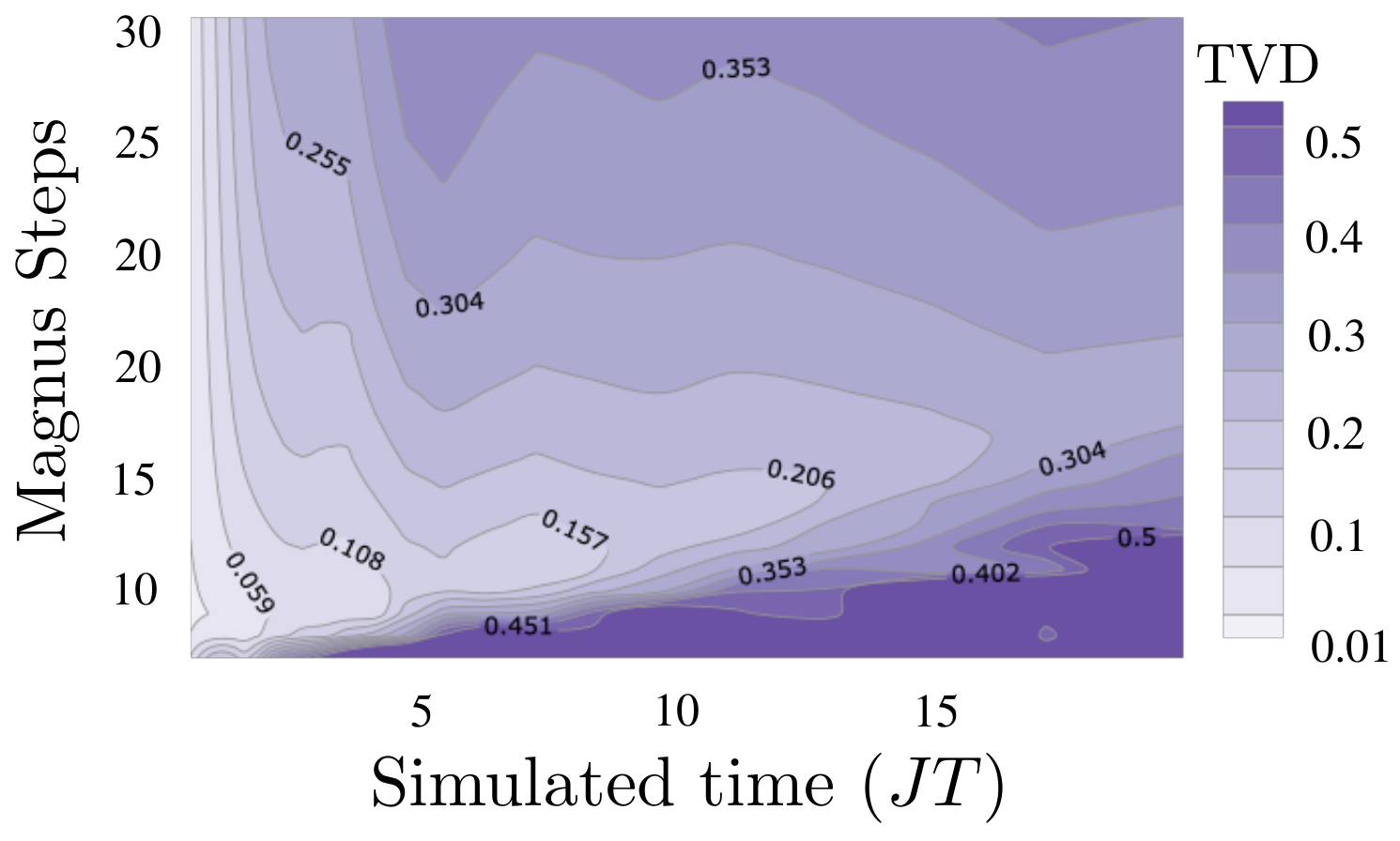} }
\caption{TVD for the phase damping noise model (noisy discretized 1) for different choices of the number of Magnus steps and simulation time of $JT$. The number of Trotter steps is held fixed at 2. The bottom area of the plot has high TVD error due an insufficient number of Magnus steps for an accurate simulation while the top area has high TVD error due to the open quantum system effects. The lighter areas in the bottom-left indicates where these trade-offs yield the best results.}
\label{fig:8}
\end{figure}

One approach to mitigate the effect of noise is to reduce the number of discretization steps that are executed to find a compromise between discretization error and hardware noise. To explore this, we show in Fig.~\ref{fig:8} the empirical TVD in the presence of noise (noisy discretized 1) for different Magnus steps (we fix the number of Trotter steps to 2) and simulation times. We provide a similar analysis for the ideal case and the second noise model in Appendix~\ref{app:TVDsearch}. For a fixed $JT$, we see a clear optimum number of Magnus steps that minimizes the error, which reflects the tradeoff between discretization error and decoherence. At sufficiently large $JT$ though, the TVD is poor regardless, since the discretization is a poor approximation at a low number of Magnus steps. Thus, the noise prevents us from accessing larger $JT$, as one should expect. For the noise parameters we have chosen, this happens significantly before the adiabatic regime.

\begin{table*}
\centering
\begin{tabular}{l|c|c|c|c}
&$\text{DA}^{1.06\ JT}_{2,2} $ &$\text{DA}^{3.40\ JT}_{3,2} $  & $\text{DA}^{5.26\ JT}_{4,2} $ & $\text{DA}^{100\ JT}_{70,2} $     \\
\hline
\hline
TVD Ideal    & 0.066 & 0.045 & 0.050 & 0.010 \\
\hline
TVD Noisy (1)   & 0.039  & 0.073 &  0.156 & 0.587    \\
\hline
TVD Noisy (2)  &  0.083 & 0.251 &  0.374 & 0.643   \\
\hline
TVD NISQ HW  & 0.280 & 0.679 & 0.596  & 0.497   \\
\hline
TVD New HW   & 0.242 & 0.268 & 0.281 & 0.639  \\
\hline
TVD New HW with DD `$XYXY$'   & 0.125  & 0.172 &  0.380  & 0.670 \\ \hline
TVD New HW with DD $\text{UR}_{10}$   & 0.127  & 0.238 &  0.320 & 0.643   \\
\end{tabular}
\caption{TVD values for the distribution for different vales of $JT$ for different computing assumptions: ideal case, the two different noisy simulations, \texttt{ibmq\_mumbai} (IBM) (NISQ HW), and the newer \texttt{ibmq\_osaka} 1.1.8 without dynamical decoupling (New HW) and with two different dynamical decoupling sequences (New HW with DD), `$XYXY$' and $\text{UR}_{10}$ \cite{DD}.  Each was simulated with $N_M$ Magnus steps and $N_T$ Trotter steps ($\text{DA}_{N_M,N_T}^{JT}$). The number of Magnus steps is chosen to correspond to the minima from Fig.~\ref{fig:8}. For the two open system simulations, Noisy (1) only includes dephasing errors, and Noisy (2) uses the \texttt{ibmq\_mumbai} (IBM) Mock backend). For the NISQ HW, the $1\sigma$ error bar is 0.005. Mock backed noise version: 1.4.5. \cite{mumbai_qiskit_mock} Real mumbai 1.10.0. \cite{mumbai_qiskit_real}. The lower TVD for Noise (1) relative to the ideal case is discussed in Appendix~\ref{apx:SVMC}.}
\label{tbl:h4t-tvd}
\end{table*}

Finally, we provide a sample of results for reduced accuracy requirements in Table \ref{tbl:h4t-tvd}, which compares the TVD values of different computing models, including demonstrations on two generations of physical hardware, for different $N_M, N_T, JT$. The number of Magnus steps is chosen to minimize the TVD for the noisy simulators. We have also included the impact of applying dynamical decoupling to our circuits on the new generation of hardware. In particular we have implemented two sequences of pulses, the `$XYXY$' sequence and the $\text{UR}_{10}$ sequence \cite{DD} on the 127 qubit device \textit{ibm\_osaka} (1.1.8).

We see from this sample of results that a TVD of $<0.1$ is achievable for sufficiently small $JT \leq 1$ by our noisy simulators but not by the hardware. While far from the adiabatic limit, $JT= 1$ is approximately where we observe the first splitting from the uniform probability for the Ising ground states (see Fig.~\ref{fig:HT4_GSs_Populations}). We can interpret this simulation time regime as when the unitary operator (Eq.~\eqref{eq:Usolution}) starts to deviate noticeably from the identity, and hence when we first begin to noticeably see the imprint of the Ising Hamiltonian. However, we should be careful about over-interpreting this apparent success for the noisy simulators because the TVD is not a good indicator of state fidelity here. The fidelity for the ideal discretized simulation is approximately $0.99$, while it is approximately $0.80$ for the dephasing only noise model; we discuss this further in Appendix \ref{apx:SVMC}. Nevertheless, if useful information can be gleaned from the state populations in this simulation time regime, then it may be accessible with less noisy hardware.

This sample of results also shows that achieving a TVD less than 0.1 is already impossible for $JT>5$ for the noise models we consider. The results for the old generation physical hardware are significantly worse than both our noisy simulators, exhibiting a TVD greater than 0.25 already at $JT = 1.06$. The TVD of the hardware clearly does not match the noise model meant to emulate it, exhibiting a faster increase in TVD for increasing $JT$ initially but also a decrease in TVD for increasing large $JT$. We leave understanding this discrepancy between the open system model and the hardware performance for future work.

Comparing the two generations of hardware, we observe that the new generation of IBM devices has improved in terms of noise due to the better performance of the raw circuits. Using dynamical decoupling further improves performance for short $J T$, even with the extra errors introduced by the single qubit gates of the sequence. For the case $JT = 1.06$ the TVD is halved for both dynamical decoupling sequences. At $JT = 3.40$ the $\text{UR}_{10}$ sequence breaks even, while the `$XYXY$' sequence can still slightly improve the performance. Finally from $JT = 5.26$ both sequences are not able to compensate for the error introduced by the additional gates.

\section{Comparison to Analog Quantum Annealing} \label{sec:analog}

We now consider the cost of simulation using an analog quantum computer. For our purposes, an analog quantum computer is defined by an initial state $\ket{\Psi(0)}$ and a family of time-varying Hamiltonians $\hat{H}^{(A)}(s)$, which can be implemented by the analog hardware platform. (We note this is not as general as the definition in Ref.~\cite{Deutsch2020}.) The execution of the analog quantum computer for time $T$ consists of evolving the  Schr\"odinger equation for $s \in [0,1]$ with $\hat{H}^{(A)}(s)$. 
From this definition it is clear that analog quantum computing devices are particularly well suited for simulating time-varying Hamiltonians, with the primary challenge being how well the computing hardware's Hamiltonian $\hat{H}^{(A)}$ can emulate the application Hamiltonian $\hat{H}^{(P)}$. The required accuracy of emulation is expected to vary depending on what is being simulated (some simulation tasks may be more robust to errors than others) \cite{Hauke2012} but also expected to grow with system size~\cite{Zhu2016,Alb2019} and longer simulation times. This makes scaling analog hardware to tackle relevant application problems a significant challenge.

As a first step in our analysis, we consider $\hat{H}^{(A)}= H_0 \hat{H}^{(P)}$, where $H_0$ is a scale factor that allows us to set the overall energy scale. We emphasize the energy scale set by the analog hardware because it directly impacts the physical runtime of the simulation. 
A higher energy scale reduces the physical runtime, which is advantageous in that it reduces the effect of decoherence since we are reducing the time the system interacts with its environment. In order to demonstrate this, we consider two different open system models for our analog quantum computer: (1) the weak coupling limit via an adiabatic master equation (AME) \cite{Davies1974,ame}, and (2) the singular coupling limit (SCL) master equation \cite{Breuer2002}, where in both cases we take $H_0 B(1) J= 1$  rad/ns and the system-bath coupling to give a single qubit dephasing time in the computational basis of $100$ns.  Further details of these two open system models are given in Appendix \ref{sec:analog-open-system}. We make this choice because of recent results from D-Wave showing that their devices can implement coherent evolution on this timescale \cite{King2022}. The key difference between the two open system models is the basis in which the dephasing happens.  For the SCL, the dephasing is in the computational basis, whereas for the AME, the dephasing is in the instantaneous energy eigenbasis.  This makes the AME particularly innocuous to adiabatic evolution since the loss of coherence between energy eigenstates is not as detrimental for the success of the evolution.  However, the AME does include thermalization to the Gibbs state of the Hamiltonian $\hat{H}^{(A)}$, so very long evolutions at high temperatures can be detrimental.

\begin{figure}[t!]
    \centering
    \subfigure[]{\includegraphics[width=0.455\textwidth]{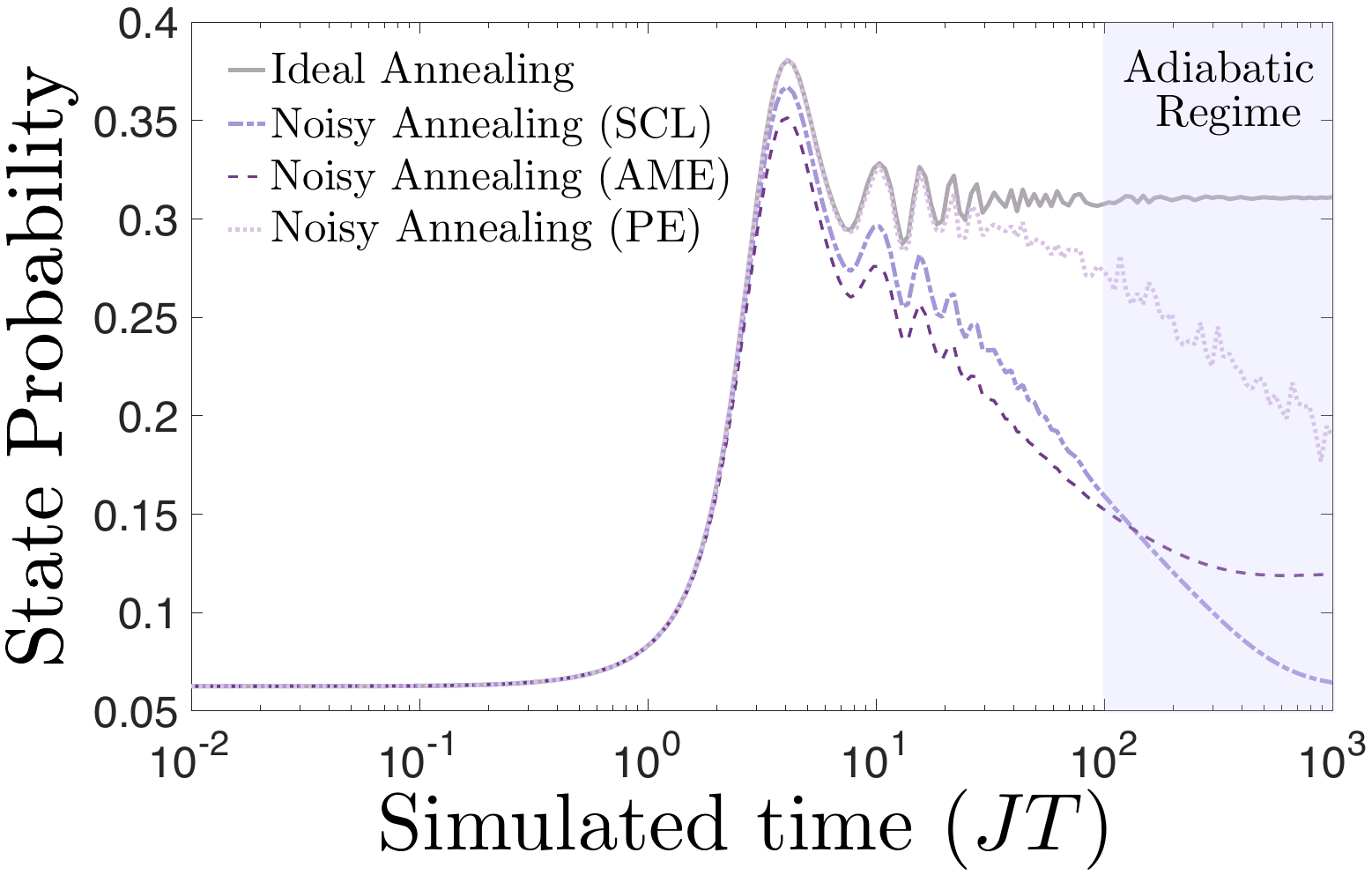} \label{fig:9a}}
    \subfigure[]{\includegraphics[width=0.45\textwidth]{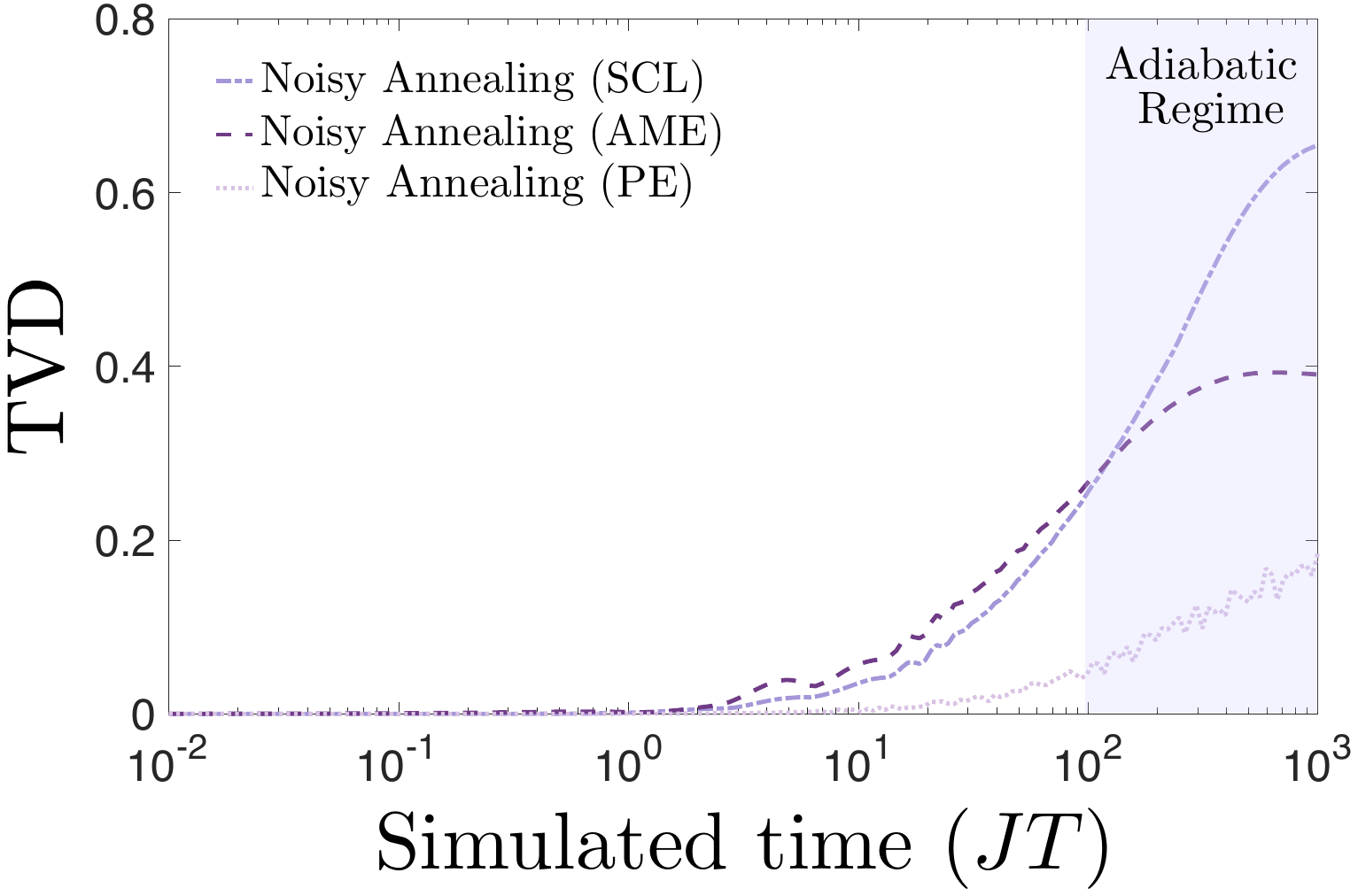} \label{fig:9b}}
    \caption{
An analysis of the impact of open system effects and programming errors on simulation quality as a function of simulation time $J T$ for analog quantum computers.  Results for the three error models discussed in the text are shown: the singular coupling limit (SCL), the adiabatic master equation (AME), and isolated programming error (PE) on the Ising Hamiltonian. (a) The probability of a specific ground state signature from $\hat{H}_{\mathrm{T}4}$, which should converge to a state probability of 1/3 in the large $J T$ limit.
(b) The TVD values for the full state distribution produced. For the SCL and AME, system-bath couplings are chosen so that the dephasing time as measured at $s=1$ is $100$ ns. Further details are provided in Appendix~\ref{sec:analog-open-system}.
These results identify open quantum system effects as the primary driver of solution quality degradation in analog quantum annealing, for the hardware-motivated parameters considered in this work.} \label{fig:Open2}
\end{figure}

The results of the simulations are shown in Fig.~\ref{fig:Open2}. Comparing against Fig.~\ref{fig:Open}, we see that the AME and SCL for our choice of parameters are able to more closely follow the ideal evolution for longer $JT$ than the circuit-model simulations, even if the coherence time is significantly shorter (100ns vs 100$\mu$s). For example, the TVD reaches 0.1 at around $JT = 20$ as opposed to around $JT = 4$ for our purely dephasing noise model (noisy discretized 1). This is a consequence of the physical runtime of the simulations on the analog hardware being significantly shorter.

The performance of the AME depends strongly on the temperature used.  As we show in Appendix~\ref{sec:analog-open-system}, reducing the temperature from $15$ to $2.38$ mK while holding the dephasing time at $100$ ns gives better performance for the AME at large simulation times.  This is primarily because the thermal state throughout the anneal has a larger weight on the ground state, which suppresses excitations out of the ground state at the early and mid stages of the evolution. This is in spite of a larger system-bath coupling to keep the dephasing time fixed.

In the analysis so far, we have not discussed the effects of implementation errors, even though these are expected to be a challenge for scaling the performance of analog hardware~\cite{Zhu2016,Alb2019}. We study the effect of implementation or programming errors (PE) by adding Gaussian noise with mean 0 and standard deviation $\sigma = 0.03 \ \mathrm{rad}/ns$ to the Ising parameters \cite{nelson2021single}.
We see in Fig.~\ref{fig:Open2} that the impact of this noise only begins to manifest at later times; for example the TVD reaches 0.1 only at approximately $JT = 200$.  
Consequently, for the parameter regimes we are considering, dephasing and relaxation effects, not programming errors, are the primary source of solution quality degradation.
\begin{figure}[t!]
    \centering
    {\includegraphics[width=0.5\textwidth]{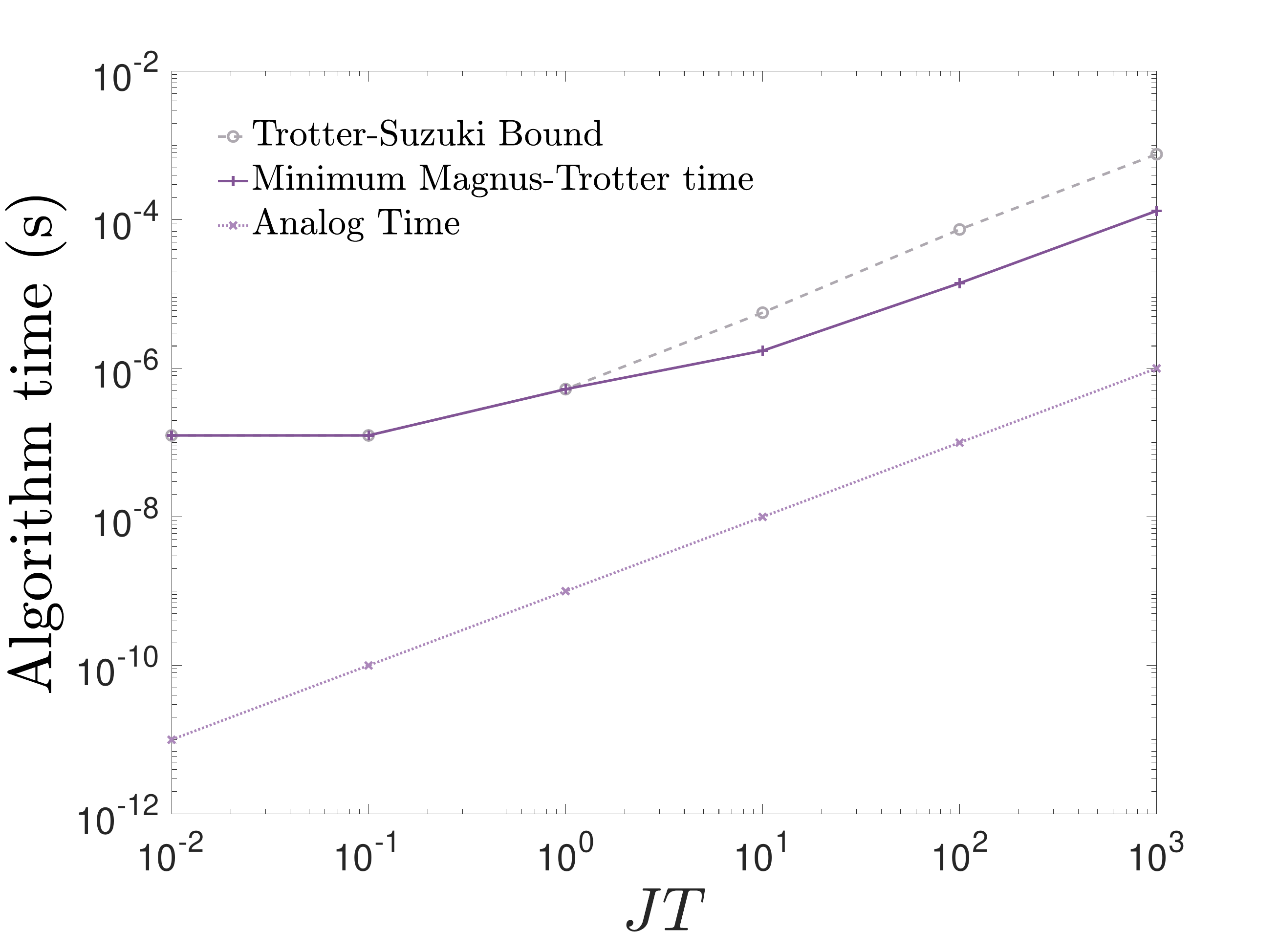}}
    \caption{Estimated algorithm runtimes for simulating the $\hat{H}_{\mathrm{T}4}$ model to a TVD of $10^{-2}$ on an ideal circuit-model quantum computer using our optimized discretization scheme (denoted ``Minimum Magnus-Trotter time''), using the same Magnus steps as our optimum but the Trotter-Suzuki bound for the number of Trotter steps, and assuming perfect rotation implementation and a gate-time of $25$ ns. The analog time corresponds to a physical implementation of the Hamiltonian with an energy scale of $B(1) J = \text{rad}/ns$ (reducing this energy scale raises the curve accordingly). Under these assumptions, an ideal analog quantum computer is two orders of magnitude faster than the ideal circuit-model quantum computer.} \label{fig:image1} 
\end{figure}

\section{Runtime Estimates}

As a final comparison between the discretized and the analog simulations, we estimate the physical runtime for the two approaches to implement high quality simulations in the absence of any noise.
Observing that the quantum circuit of a single Magnus step is composed of many repetitions of the same Trotter step (shown in Fig.~\ref{fig:color_topo}), we can estimate the total physical runtime of the quantum simulation circuit by first estimating the time for the execution of one Trotter step and then multiplying by the total number of required steps (i.e., as in Table \ref{tb:ht4-mag-trot}). 
To be as hardware agnostic as possible and to give as favorable an estimate as possible, we assume that no further compilation of the 2-qubit and 1-qubit unitaries shown in Fig.~\ref{fig:color_topo} into a hardware native gate-set is needed. This means that for the single Magnus step depicted in Fig.~\ref{fig:color_topo}, we would have 4 layers of gates per Trotter step, since the last layer can be absorbed with the next Trotter step. Therefore, the total runtime of our circuit is given by
\begin{equation}
    N_M N_T \left(\chi_1 C_2  +C_1\right)+ C_1,
\end{equation}
where $\chi_1=3$ is the chromatic index for the nonzero edges in the target Hamiltonian $\hat{H}_{\text{T}4}$, $C_2$ is the runtime of implementing two qubit unitaries, and $C_1$ is the runtime of implementing single qubit unitaries.

We make the simplifying assumption that each layer of gates takes exactly $25 ns$. This assumption is based on the current execution time of single-qubit gates in superconducting architectures \cite{beverland2022assessing, Suchara_2013, PhysRevResearch.5.043194, PhysRevA.108.042617, PhysRevApplied.19.034071,L__2012}. 
Using these assumptions, the runtime of discretized simulations is shown in Fig.~\ref{fig:image1}. We see that in the intermediate and late-time limits, the actual runtime cost is almost one order of magnitude smaller than the bound in Eq.~\eqref{eq:order:trotter2}. 

For our analog simulations, we assume an energy scale of $H_0 B(1) J = 1$ rad/ns, which is approximately the energy scale in the D-Wave quantum annealing hardware based on superconducting flux qubits \cite{Joh2011}. This choice gives analog simulation runtimes that are approximately 100 times faster than our discretized simulations, as shown in Fig.~\ref{fig:image1}. 

We emphasize that these estimates of the physical runtime exclude any state preparation, device programming, and read-out times. They also assume native implementations of the arbitrary single qubit rotations and 2-qubit unitaries.  Decomposing such gates into a discrete set of universal gates \cite{Kitaev1997} would further increase the runtime.

\section{Discussion} \label{sec:discussion}
%
In this work, using optimistic values for gate runtimes we have estimated the physical runtime for accurate circuit-model simulation of a simple four-qubit quantum annealing process. The reported runtime is significant and arguably prohibitive: it is long enough that decoherence processes on physical hardware such as dephasing degrade the solution quality significantly before the adiabatic limit is well approximated. For larger problems requiring even longer annealing times to reach the adiabatic limit, we can conclude that the development of faster hardware operations, longer coherence times, and/or fault-tolerant operations will be necessary to obtain accurate simulation of these dynamics at interesting scales (i.e., with more than 100 qubits and $JT \geq 100$).

Similarly for analog simulators, the coherence time and the energy scale of the hardware set a limit on how long a simulation time $JT$ can be accurately simulated by the hardware. However, the physical runtime can be much shorter than in the discretized case, allowing analog simulators to perform accurate simulation to longer $JT$ values.  

However, implementation or programming errors are expected to be a challenge for scaling analog computations. If we have an ideal unitary (such as our time-evolution unitary) that we can only implement up to some fixed coherent error, then in the circuit model such coherent errors result in a worst-case infidelity that scales quadratically to leading order in the circuit size (both in the number of qubits and depth of the circuit) \cite{Sheldon2016,Wallman2016,Iverson2020}, and we can expect identical scaling for continuous-time computations. These conclusions apply in the worst-case, and we can expect some observable expectation values to be less sensitive than others, as illustrated in Ref.~\cite{Hauke2012}. The challenge is then to identify application problems that fit into the constraints of the analog simulator: simulation times that are low enough and a desired solution quality for specific observables that is achievable with the coherence times and programming error rates of the device.

We emphasize that the quantum annealing problem we study here is particularly simple in multiple ways.  It is only defined on four qubits, the Hamiltonian has a very simple interaction graph, and we are approximating the adiabatic limit final state well already at $JT = 100$.  Nevertheless, a high quality emulation of the final state in the adiabatic limit requires over $10^2$ Magnus-Trotter discretizations steps, resulting in a circuit depth above 400. Variational techniques could be used to optimize the circuit for a fixed number of Magnus and Trotter steps \cite{Lubasch1, Lubasch2, Andrew1}, but we do not expect the improvement to significantly change our results. While such high quality emulation is beyond current NISQ hardware, we believe that this kind of problem will serve as a useful benchmark to test the performance of future NISQ devices and early fault-tolerant hardware.

In our analysis, we have only considered Hamiltonians that can be natively embedded in the hardware connectivity graph of both the analog and circuit-based systems. If considering Hamiltonians that are not native to the analog hardware, then the analog hardware faces a critical obstacle.  While in principle perturbative gadgets \cite{Oliveira2005,Jordan2008,Cao2015,Bausch2020} can be used to implement arbitrary Hamiltonians (see Ref.~\cite{Mozgunov2023} for a recent proposal using superconducting qubits), such implementations introduce additional hardware overhead which may defeat the purpose of using the analog hardware.

While fault-tolerant hardware will allow us to perform simulations with runtimes beyond the physical decoherence time scales of the hardware \cite{Aha1997,Kni1998}, it will add additional runtime overhead to our simulation task. Consequently, one can always expect direct emulation with analog hardware to be significantly faster than simulation on fault-tolerant hardware. While fault-tolerant hardware may be the only way to access sufficiently large simulation problems with long simulation times or to perform very-high quality (low error) simulations, at intermediate scales and times analog hardware can provide a tradeoff between runtime and quality. Depending on the application, a faster but less accurate simulation may be more desirable, and analog quantum computers can fill that niche, assuming they can achieve the necessary error rates at application sizes.

Our results are not meant to represent a general benchmarking of circuit model devices and analog hardware on a broad class of problems.  Instead, our work examines the special situation where the simulation problem of interest is readily implemented by the analog hardware such that noise and overhead are minimized. The results of this work suggest that the simulation of the same quantum dynamics can require considerable resources and performance from circuit-based quantum computers. 
Consequently, it will likely be several years before such computers can reproduce the kinds of demonstrations that are currently being conducted on analog quantum computers \cite{Choi2016,Chiu2019,Koepsell2019,doi:10.1126/science.abi8794,Scholl2021,Monroe2021,King2023}.
The apparent usefulness of these devices for a small set of applications today suggests important directions for future research into the design and operation of analog quantum computers.
The utility of these machines will be determined by: (1) the types of time varying Hamiltonians they can natively support; (2) the size of the system (number of qubits), (3) the largest effective evolution time ($JT$) that they can support; and (4) how much programming errors can be suppressed or mitigated.
The opportunities and limitations for these system properties remain fairly under-explored at this time but is an active area of research~\cite{Kashyap2024}.

\begin{acknowledgments}
This material is based upon work supported by the National Science Foundation under Grant No. 2037755. J.G-C acknowledges financial support from OpenSuperQ+100 (Grant No. 101113946) of the EU Flagship on Quantum Technologies, as well as from the EU FET-Open project EPIQUS (Grant No. 899368), also from Project Grant No. PID2021-125823NA-I00 595 funded by MCIN/AEI/10.13039/501100011033 and by “ERDF A way of making Europe” and “ERDF Invest in your Future”;  from the Spanish Ministry of Economic Affairs and Digital Transformation through the QUANTUM ENIA project call - Quantum Spain, and by the EU through the Recovery, Transformation and Resilience Plan – NextGenerationEU within the framework of the Digital Spain 2026 Agenda.  J.G-C also acknowledges funding from Basque Government through Grant No.  IT1470-22 and the IKUR Strategy under the collaboration agreement between Ikerbasque Foundation and BCAM on behalf of the Department of Education of the Basque Government, as well as from UPV/EHU Ph.D. Grant No. PIF20/276.
Research presented in this article was supported by the Laboratory Directed Research and Development program of Los Alamos National Laboratory under project numbers 20210114ER, 20230546DR, 20240032DR.
This work was performed, in part, at the Center for Integrated Nanotechnologies, an Office of Science User Facility operated for the U.S. Department of Energy (DOE) Office of Science.
Sandia National Laboratories is a multi-mission laboratory managed and operated by National Technology \& Engineering Solutions of Sandia, LLC (NTESS), a wholly owned subsidiary of Honeywell International Inc., for the U.S. Department of Energy’s National Nuclear Security Administration (DOE/NNSA) under contract DE-NA0003525. This written work is authored by an employee of NTESS. The employee, not NTESS, owns the right, title and interest in and to the written work and is responsible for its contents. Any subjective views or opinions that might be expressed in the written work do not necessarily represent the views of the U.S. Government. The publisher acknowledges that the U.S. Government retains a non-exclusive, paid-up, irrevocable, world-wide license to publish or reproduce the published form of this written work or allow others to do so, for U.S. Government purposes. The DOE will provide public access to results of federally sponsored research in accordance with the DOE Public Access Plan. \url{https://www.energy.gov/downloads/doe-public-access-plan.}
\end{acknowledgments}
\appendix

\section{Shortcomings of Metrics Defined on a Single Basis}
\label{apx:SVMC}
%
\begin{table}[b!]
\centering
\begin{tabular}{c|c}
    & TVD \\
\hline
\hline 
$\text{SVMC}_{2.38 \text{mK}}$ &   0.0976 \\
$\text{SVMC}_{15 \text{mK}}$ &   0.4046 \\
$\text{DA}^{1000\ JT}_{660,2}$ & 0.0072\\

\end{tabular}
\caption{Total variation distance (TVD) values for the distribution of ground states at the adiabatic limit and the distribution obtained from the SVMC method. The parameters used for the SVMC simulation are: $\beta=3.19$, which corresponds with a temperature of $2.38$ mK, and $\beta=0.5092$, which corresponds with a temperature of $15$ mK,  if we assume that $B(1) J = 1$ GHz in the Hamiltonian. }
\label{tb:weakness}
\end{table}

In this appendix we demonstrate the weakness of the TVD metric to assess simulation quality. To do this, we give two examples of cases where a low TVD actually corresponds to a poor simulation fidelity. For the first case, we use the spin vector monte carlo (SVMC) method~\cite{svmc}. This is a classical Monte Carlo method often used to evaluate the extent to which the output statistics of quantum annealing devices can be reproduced using a classical model. This model tends to perform well for reproducing the relative populations of final ground states of the transverse field Ising Hamiltonian near the adiabatic limit, but it often fails to accurately capture the dynamics of the system since it fails to account for coherent dynamics. 
\begin{figure}[t!]
    \centering
    \subfigure[]{\includegraphics[width=0.38\textwidth]{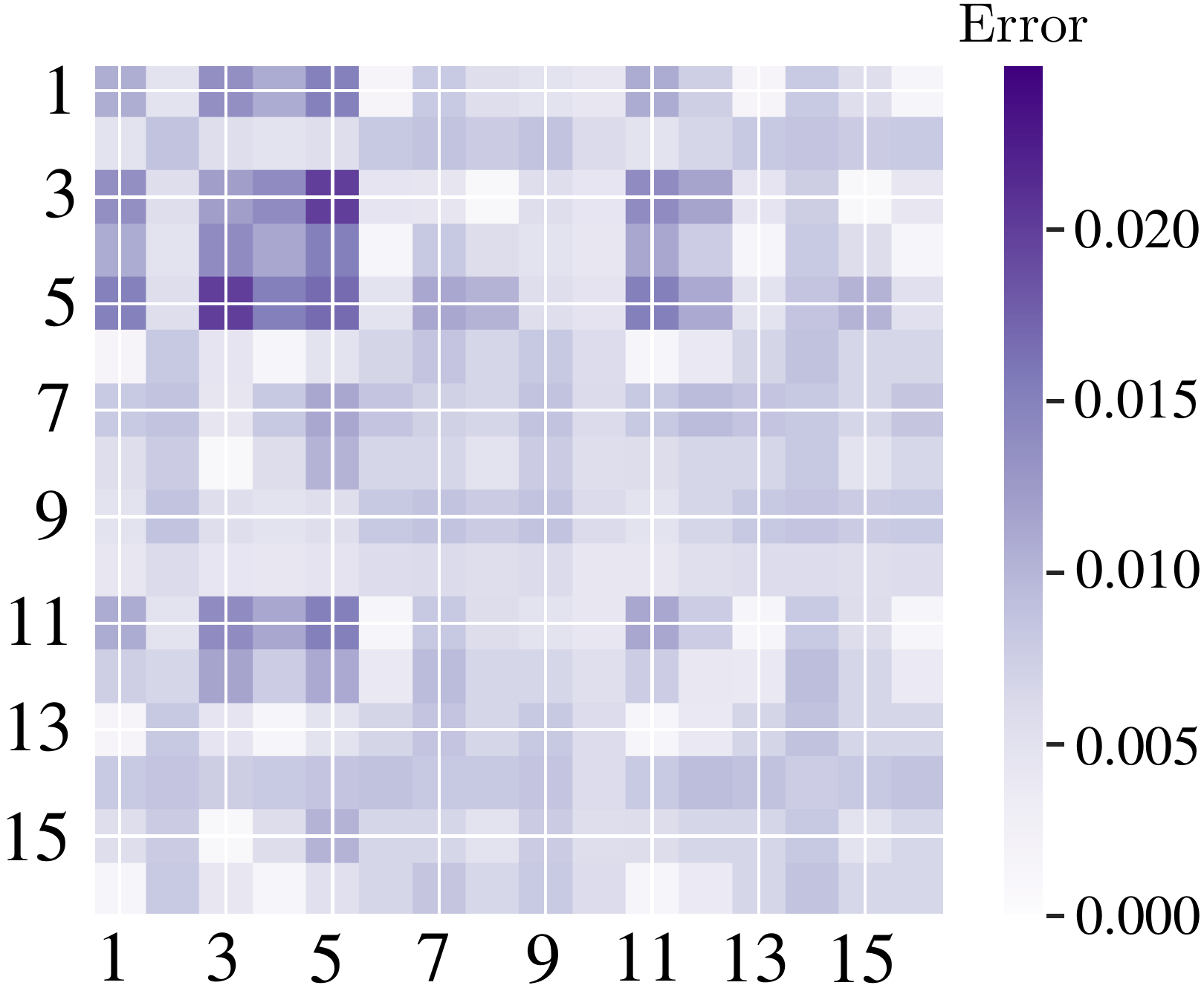}\label{fig:DensityMatrix_ideal} }
    \subfigure[]{\includegraphics[width=0.38\textwidth]{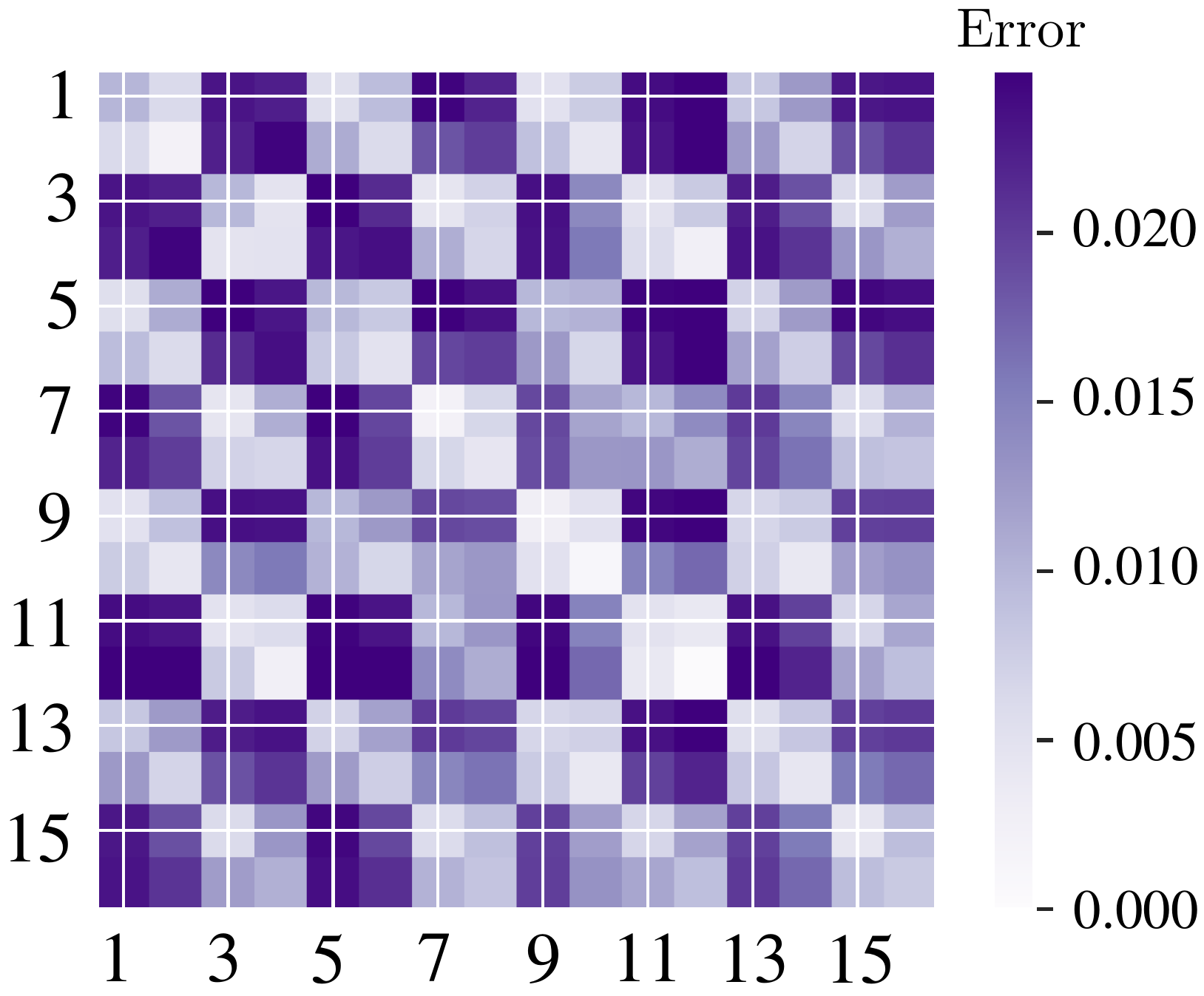}\label{fig:DensityMatrix_noisy} }
    \caption{
Absolute magnitude of the matrix components (in the computational basis) of subtracting the density matrix of the exact evolution from the density matrix of the discretized evolutions for (a) the ideal discretization and (b) the Noisy discretized 1 simulator.}\label{fig:DensityMatrix}
\end{figure}

The model simulates the transverse field Ising Hamiltonian by modeling the spins as classical rotors in the $x$-$z$ plane. The Hamiltonian for the rotors is taken to be 
\begin{eqnarray}
{H}(\theta) &=& -A(t)\sum_i{\sin(\theta_i)} \nonumber \\
    && +B(t) \left(\sum_{\langle i,j \rangle} J_{ij} \cos(\theta_i)\cos(\theta_j) +\right. \nonumber \\
    &&  \hspace{2.5cm} \left .\sum_i h_i {\cos(\theta_i)} \right) ,
\end{eqnarray}
where the $i$-th spin is replaced by a unit length rotor pointing along the direction given by  $\theta_i$.  Monte Carlo updates of the rotors are performed along the annealing schedule given by $A(s)$ and $B(s)$. For each Monte Carlo sweep, the direction of each spin is updated to a new random angle $\theta_i' \in [0,2\pi)$ with probability given by a Metropolis-type update, $\min({1, \exp{-\beta \Delta E_i}})$ where $\beta$ is the inverse temperature and
\begin{eqnarray}
\Delta E_i &=& -A(s) [\sin(\theta_i') - \sin(\theta_i)]   + B(s) [ h_i (\cos{\theta_i'} - \cos{\theta_i}) \nonumber \\
    &&  +  \sum_{j | j\neq i} J_{ij} \cos{\theta_j} (\cos{\theta_i'} - \cos{\theta_i})]
\end{eqnarray}
At the end of the anneal, rotors with $\cos(\theta_i) > 0$ are mapped to spin up, and rotors with $\cos(\theta_i) < 0$ are mapped to spin down.  This gives a mapping from the rotor orientation to computational basis measurement outcomes.

For each parameter choice, we performed 1,000 independent trials of SVMC.  Each trial used 10,001 Metropolis updates, where $s$ is evenly spaced and updated in ascending order with values $\{0, 0.0001, \dots, 1\}$.  Empirically, these parameters were sufficient to ensure convergence to the steady state distribution with high probability.

We calculate the TVD of the distribution of the computational basis states obtained from the SVMC simulations and the theoretical adiabatic limit result, and the results are given in Table~\ref{tb:weakness}. Even though SVMC finds the Ising ground states with high probability, at high temperatures (15 mK) the population distribution does not closely match the closed quantum system distribution. However, as the temperature is lowered, for example to 2.38 mK, the distribution is below 0.1 TVD.  We expect that fine tuning of the SVMC simulation parameters can further improve this result. Therefore, while SVMC does not simulate coherent states, it is able to generate distributions that score well on the TVD metric because it can reproduce the classical probability distribution associated with computational basis measurements. This result highlights the primary weakness of the TVD metric, but it is not an issue for the results in the main text since we know that the approaches we study there that have a low TVD metric also have a low infidelity.  

Another example of the weakness of the TVD metric and how it fails to capture coherences is illustrated in Table~\ref{tbl:h4t-tvd} in the main text.  For the smallest $JT$ value considered, we observe a smaller TVD for one of the noisy simulators (noisy discretized 1) than for the ideal discretization. However, a calculation of the fidelity shows that the ideal case has a significantly larger fidelity (0.99 versus 0.80). To show why this is happening, we consider the components of the matrix calculated by subtracting the density matrix of the exact evolution from the density matrix of the discretized evolutions in the computational basis. We show in Fig.~\ref{fig:DensityMatrix} a visualization of the absolute value of the components of these matrices. What we find is that while the off-diagonal elements differ considerably for the noisy case, the diagonal elements are actually closer to the ideal results than the ideal discretized case.  This explains why the TVD reported in Table~\ref{tbl:h4t-tvd} is lower, even though the state has a much lower fidelity.

\section{Empirical TVD Search }\label{app:TVDsearch}

\begin{figure}[t!]
    \centering
    \subfigure[$JT = 10$]{\includegraphics[width=0.48\textwidth]{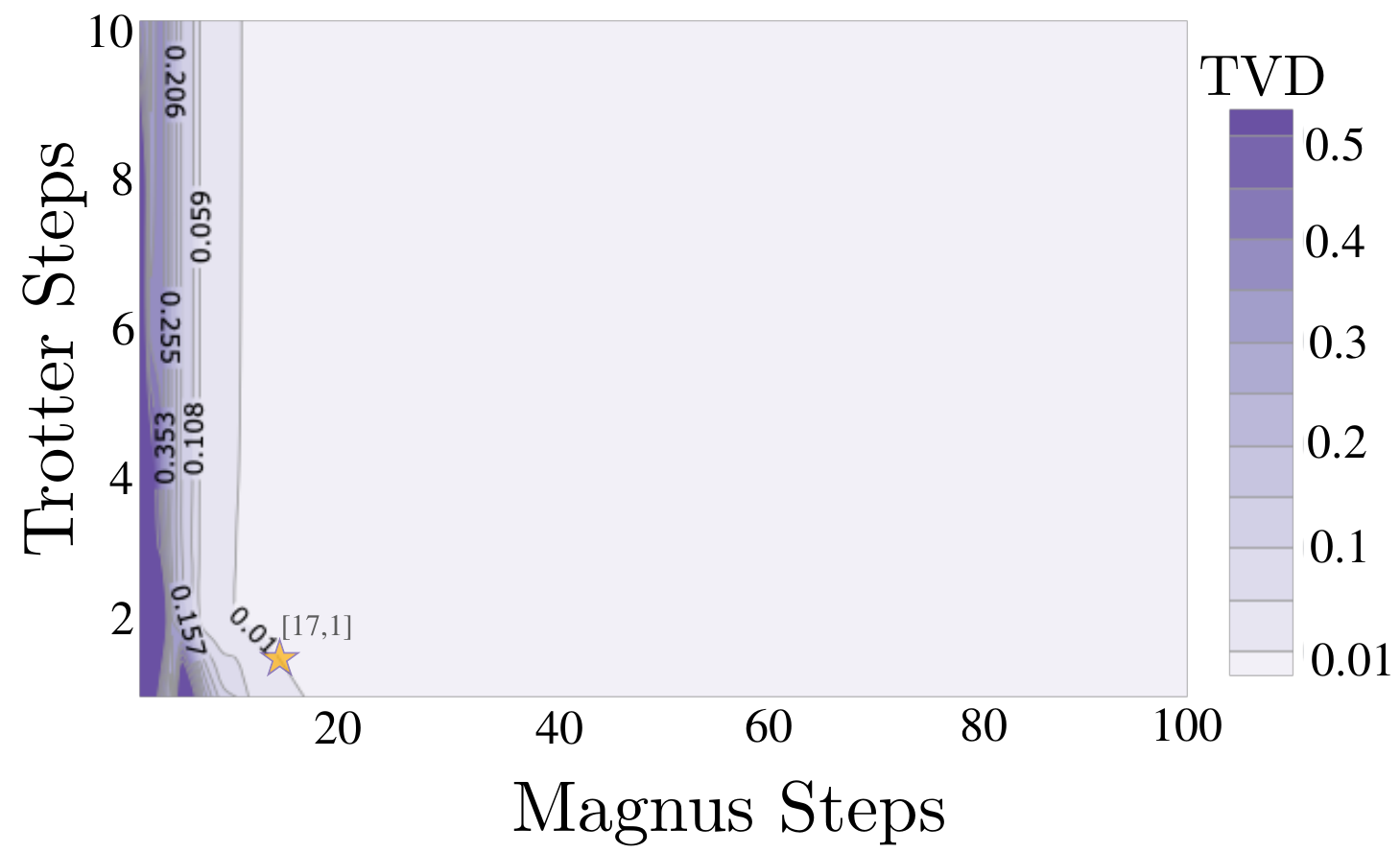} 
    \label{fig:5_Appendix_10}}
    \subfigure[$JT = 10^3$]{\includegraphics[width=0.48\textwidth]{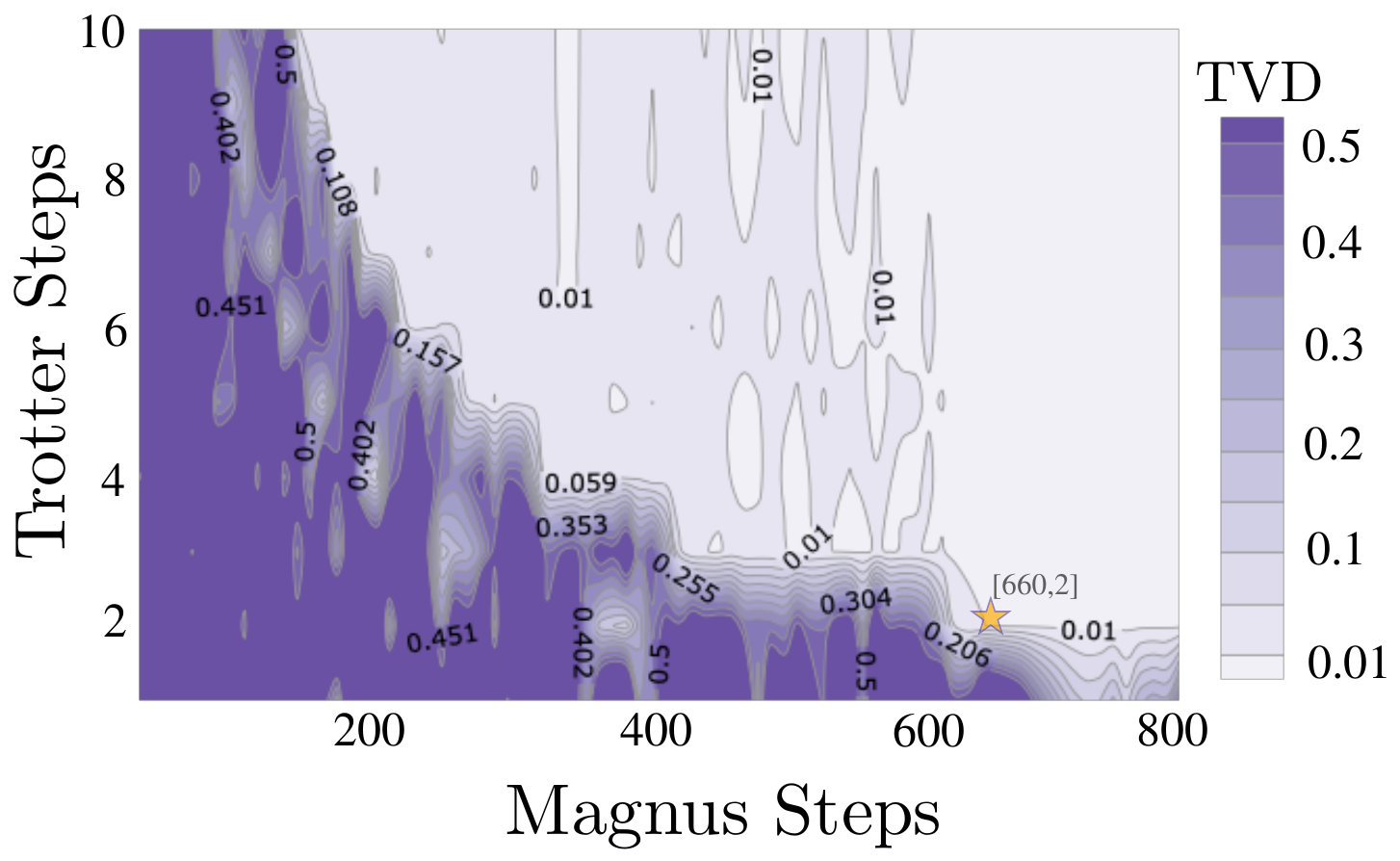} 
    \label{fig:5_Appendix_1000}}
    \caption{TVD for different choices of the number of Magnus and Trotter steps for a simulation time of (a) $JT = 10$ and (b) $JT = 10^3$. The ``star'' corresponds to a choice that achieves a TVD $<10^{-2}$ and minimizes $N_M N_T$.  It corresponds to the value reported in Table~\ref{tb:ht4-mag-trot}.}\label{fig:Open3}
\end{figure}

We give additional examples of parameter search results in Fig.~\ref{fig:Open3} to complement the result presented in the main text in Fig.~\ref{fig:TVD_Search}.

\begin{figure}[tb!]
    \centering
    \subfigure[Ideal Discretized]{\includegraphics[width=0.5\textwidth]{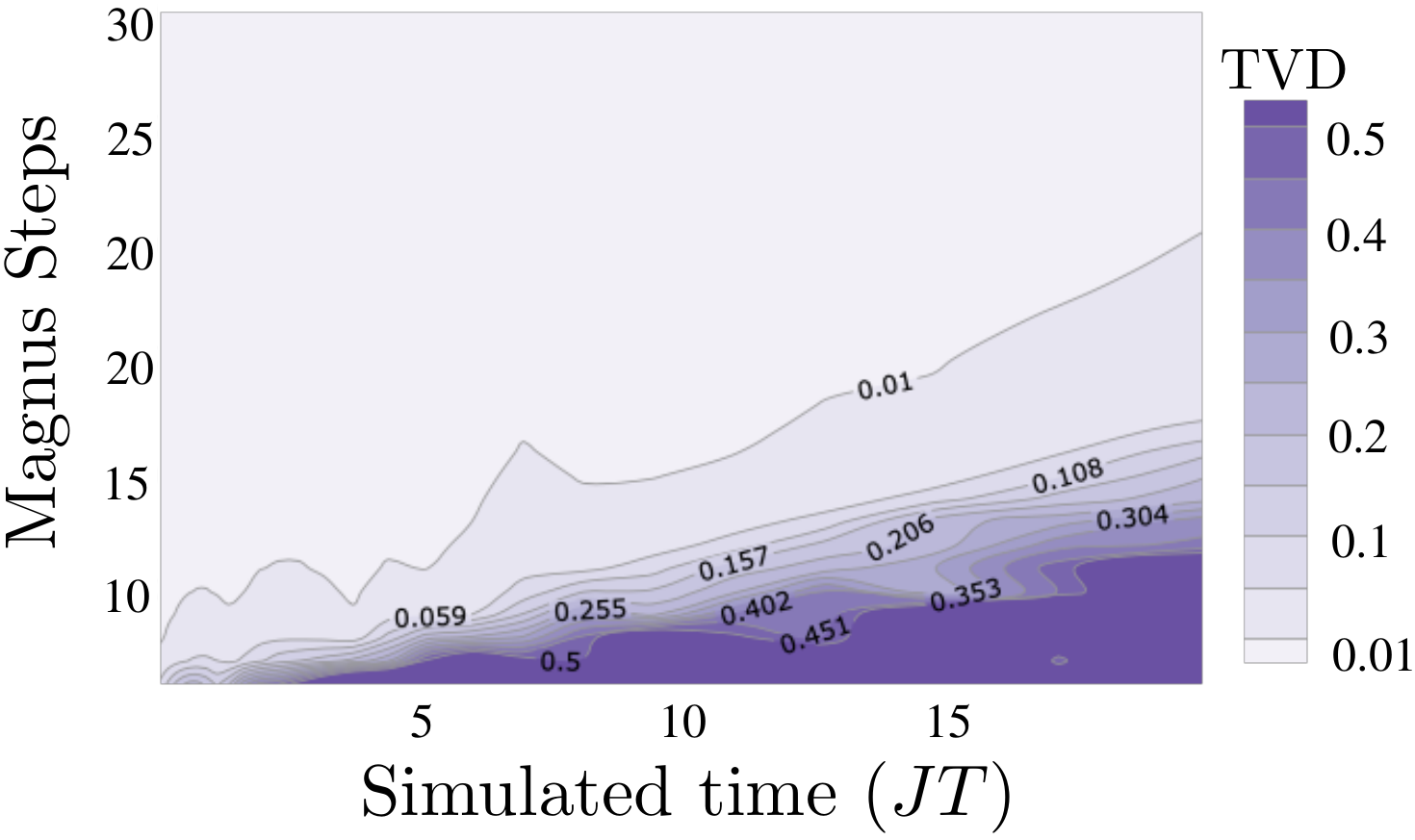} \label{fig:AppendixB_a}}
    \subfigure[Noisy Discretized 2]{\includegraphics[width=0.495\textwidth]{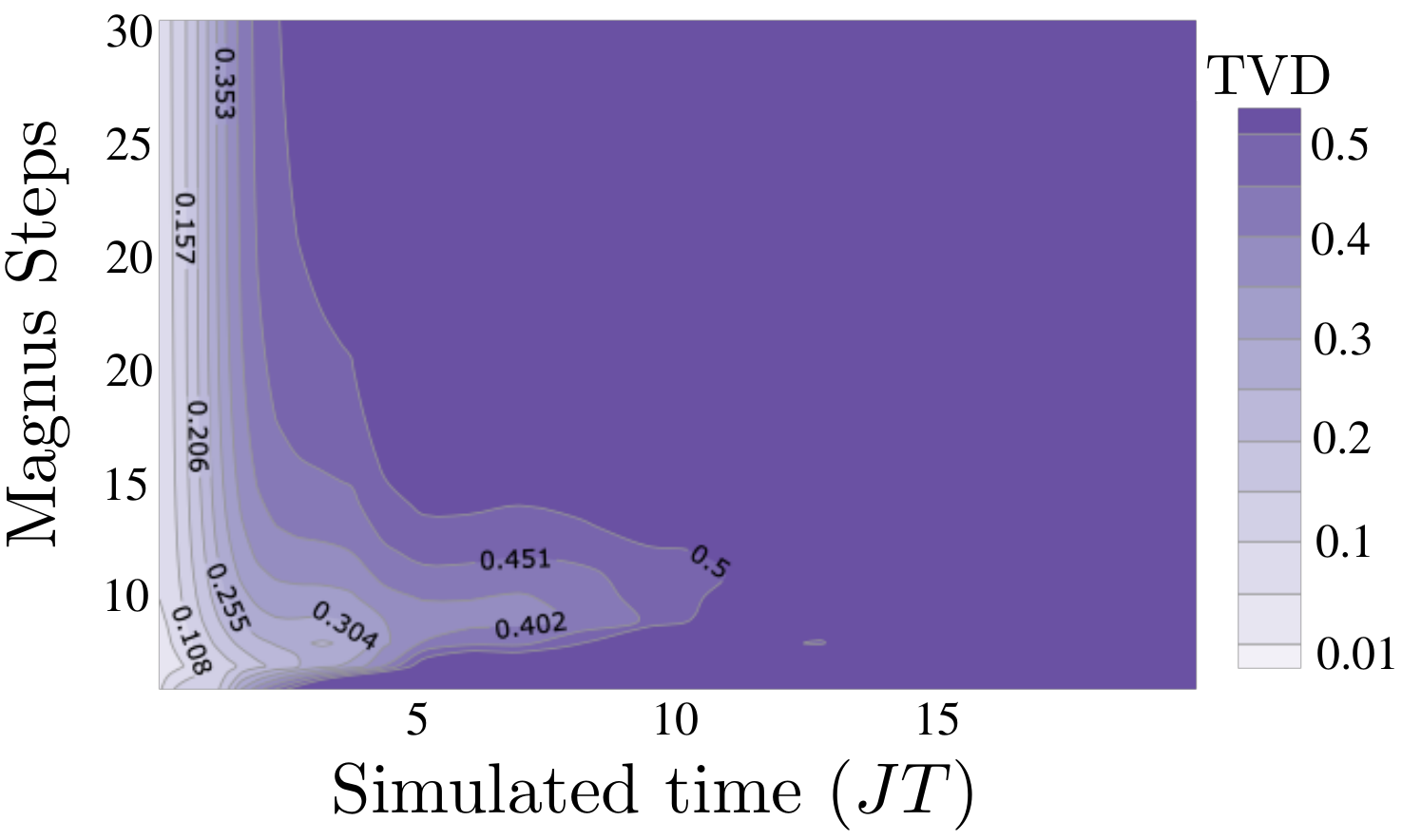} \label{fig:AppendixB_b}}
    \caption{TVD for (a) the ideal model and (b) the second noise model (noisy discretized 2) for different choices of the number of Magnus steps and simulation time of $JT$.  The number of Trotter steps is held fixed at 2.}\label{fig:Open4}
\end{figure}

We also give additional examples of parameter search results in Fig.~\ref{fig:Open4} to complement the result presented in the main text in Fig.~\ref{fig:8}. The second noise model [Fig.~\ref{fig:AppendixB_b}] also exhibits a minimum in the TVD for a fixed $JT$, but the TVD in general is worse than the simpler noise model presented in the main text.  This is to be contrasted with the ideal case [Fig.~\ref{fig:AppendixB_a}], where increasing the Magnus step improves the TVD.

\begin{figure}[tb!]
    \centering
    {\includegraphics[width=.485\textwidth]{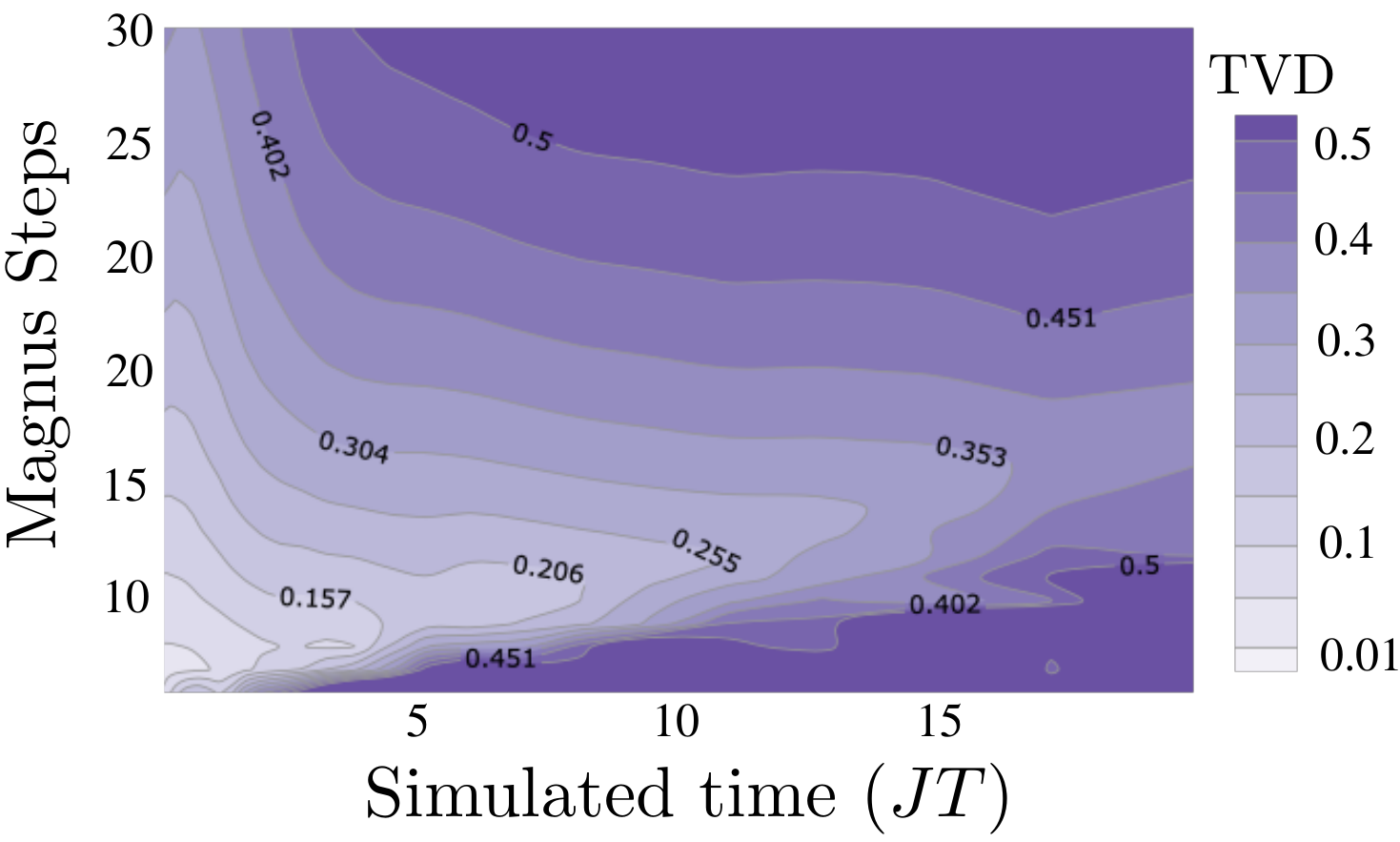} }
\caption{TVD for the phase damping and thermal relaxation noise model (noisy discretized 3) for different choices of the number of Magnus steps and simulation time of $JT$. The number of Trotter steps is held fixed at 2. The bottom area of the plot has high TVD error due an insufficient number of Magnus steps for an accurate simulation while the top area has high TVD error due to the open quantum system effects. The lighter areas in the bottom-left indicates where these trade-offs yield the best results.}
\label{fig:8Appendix}
\end{figure}

\begin{figure}[tb!]
    \centering
    \subfigure[]{\includegraphics[width=0.475\textwidth]{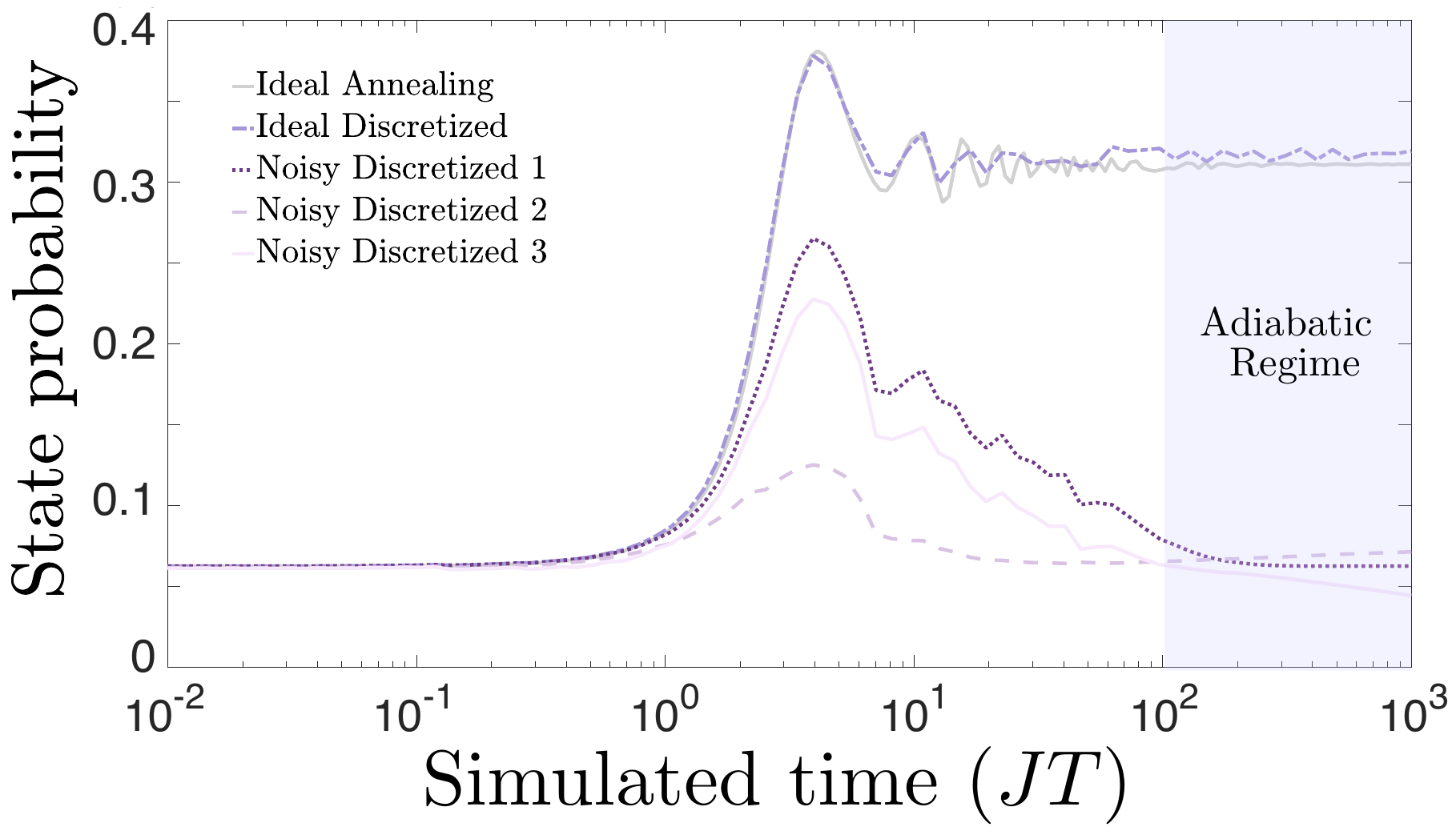} \label{fig:StateProbabilityOpenAppendix}}
    \subfigure[]{\includegraphics[width=0.485\textwidth]{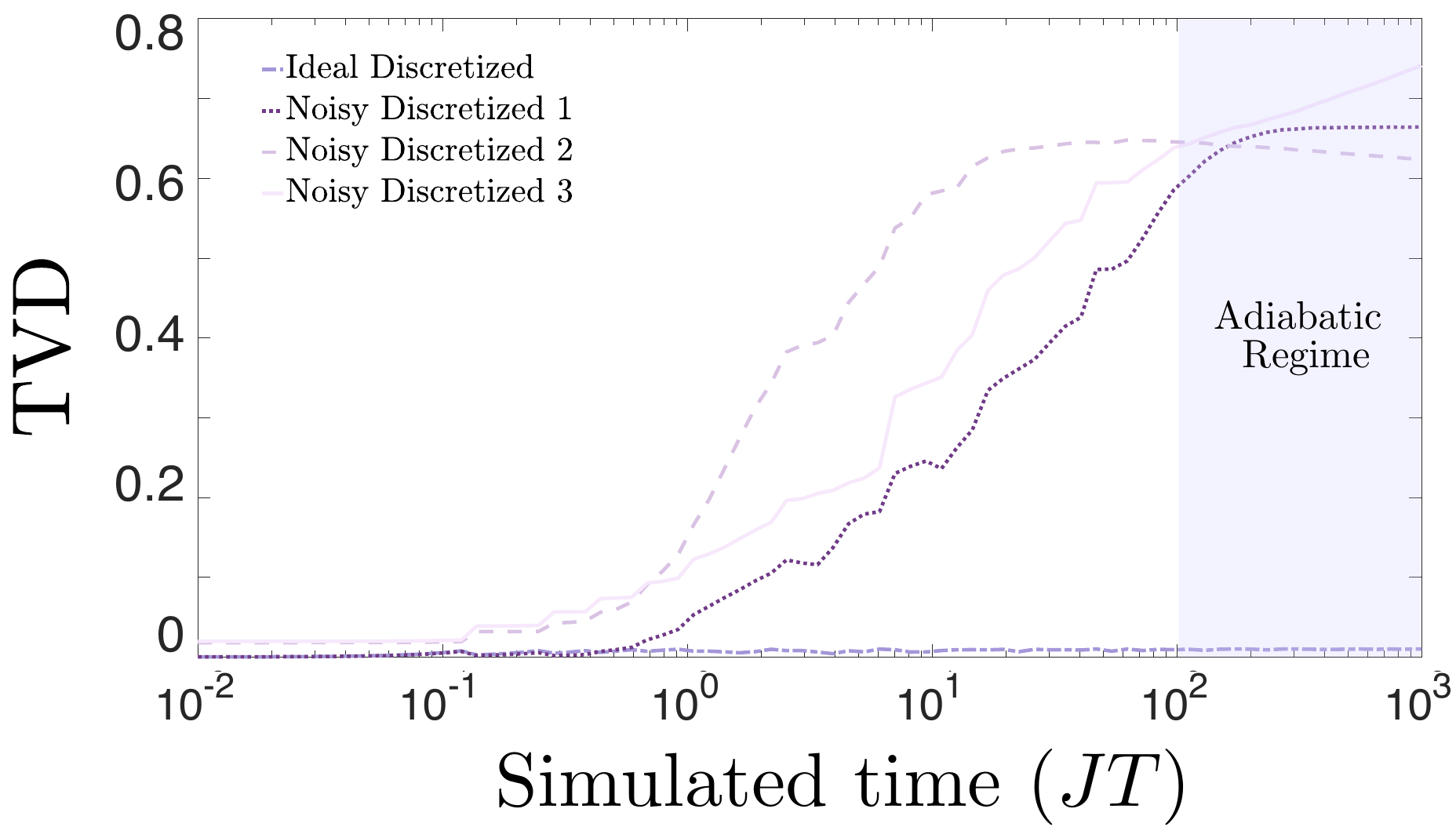} \label{fig:TVDOpenAppendix}}
    \caption{
An analysis of the impact of open system effects on the simulation quality as a function of simulation time $J T$. Ideal Discretized corresponds to simulations with  $N_T = 2$ and the minimum number of Magnus steps $N_M$ needed to achieve a TVD $< 0.01$ in the ideal discretized case. Noisy Discretized 1 are open system simulations with only phase damping as described in the main text. Noisy Discretized 2 are open system simulations with readout
error, depolarizing error and thermal relaxation error as described in the main text. Noisy Discretized 3 are open system simulations with phase damping and thermal relaxation only. (a) Probability of the ground state $\ket{\uparrow \uparrow \downarrow \uparrow }$ signature from $\hat{H}_{\mathrm{T}4}$, which should converge to a state probability of 1/3 in the long-time limit.
(b) TVD for the full state distribution produced and includes a uniform distribution as a point of reference. }\label{fig:OpenAppendix}
\end{figure}

In Figs.~\ref{fig:8Appendix} and~\ref{fig:OpenAppendix}, we provide additional open system simulation results for a third open system for the discretized simulations.  This third model (noisy discretized 3) includes phase damping and thermal relaxation errors only, characterized now by both the $T_2$ and the $T_1$ time respectively, in contrast to the Noisy discretized 1 model in the main text that only includes phase damping.  Details of the parameter choices for this model are given in Appendix \ref{apx:lima}. This model does slightly worse in terms of the TVD compared to the Noisy Discretized 1 noise model, but the similarity of the results suggest that the dominant source of error is dephasing.

\section{Error bounds in Trotterization}
\label{apx:error_bounds}
Let us consider two quantum states $\hat{\rho}$ and $\hat{\sigma}$ and the probability distributions induced by measuring both of them in the computational basis.
Let us call the probability distributions $p_{\rho}$ and $p_{\sigma}$ 
respectively. Then, the total variation distance between $p_{\rho}$ and $p_{\sigma}$ 
satisfies
\begin{equation}
    \text{TVD}(p_{\rho},p_{\sigma})\leq \lvert\lvert \hat{\rho} -\hat{\sigma} \rvert\rvert_{T} ,
\end{equation}
with $||\cdot||_T = \frac{1}{2} \lVert \ \cdot \  \rVert_1$ denoting the trace distance. Therefore, if we assume that $\hat{\rho}=\hat{U}_1 \hat{\rho}_0 \hat{U}_1^\dagger$ and $\hat{\sigma}= \hat{U}_2 \hat{\rho}_0 \hat{U}_2^\dagger$, then we have the inequality $\lvert\lvert \hat{\rho}-\hat{\sigma}\rvert\rvert_{T} \leq \lvert\lvert \hat{U}_1-\hat{U}_2\rvert\rvert_\infty $ with $||\cdot||_\infty$ the operator norm. We use this result to give an upper-bound for the TVD between the exactly evolved state and the state evolved using our discretized evolution
\begin{equation}
    \text{TVD}(p_\text{exact},p_\text{appr})\leq \lvert\lvert \hat{U}_\text{exact}-\hat{U}_\text{appr}\rvert\rvert_{\infty} . 
\end{equation}

We define $\epsilon:=\lvert\lvert \hat{U}_\text{exact}-\hat{U}_\text{appr}\rvert\rvert_{\infty}$  and use Eq.~(\ref{eq:order:trotter2}) for the approximation of the Trotter error of $\prod_{i=1}^{N_M} e^{\hat{A}_i+\hat{B}_i}$, such that the total number of steps $N_M N_T\sim $ ${\sim \sum_{i=1}^{N_M} \Big\lceil \sqrt{\frac{(T/N_M)^3 (||[[\hat{A}_i,\hat{B}_i],\hat{B}_i]||+0.5||[[\hat{A}_i,\hat{B}_i],\hat{A}_i]|| }{12\epsilon}}\Big\rceil} $. If we apply this formula with $\epsilon \sim 0.01$ to our Hamiltonian in Eq.~\eqref{eq:T_sig} and we assume each layer of gates takes $25$ ns, then the run time given by the theoretical bound is 
\begin{multline}
\footnotesize
    \mathrm{T}\sim  25 \sum_{i=1}^{N_M} \left(\frac{100\left(\frac{T}{N_M}\right)^3 (||[[\hat{A}_i,\hat{B}_i],\hat{B}_i]||)}{12}\right.\\ \left. +\frac{0.5||[[\hat{A}_i,\hat{B}_i],\hat{A}_i]||}{12}\right)^{0.5} , 
\end{multline}
with T in units of ns. For our particular Hamiltonian, we have
$$\hat{A}_i=\int_{(i-1) \delta}^{i \delta} ds (1-s) \hat{H}_0= (\delta+\frac{\delta^2}{2}-i\delta^2)\hat{H}_0,$$
and
$$\hat{B}_i= \int_{(i-1) \delta}^{i \delta} ds \ s \hat{H}_\text{Target}= \frac{\delta^2}{2}(2i-1)\hat{H}_\text{Target},$$
where $\delta = 1/N_M$, $i=1\ \dots N_M$. Additionally in our particular case %
\begin{eqnarray}
\left\| \left[ \left[ \hat{H}_0, \hat{H}_{\text{Target}} \right], \hat{H}_{\text{Target}} \right] \right\| + \\ 0.5 \left\| \left[ \left[ \hat{H}_0, \hat{H}_{\text{Target}} \right], \hat{H}_0 \right] \right\|_{\infty} \nonumber \\
= 54.7660. \nonumber
\end{eqnarray}

\section{Leading approximation error between Trotter and Magnus expansion}
\label{apx:laTM}

In this appendix, we sketch how the Magnus approximation error
and the Suzuki-Trotter product formula error compete with each other
in solving some time dependent Schrodinger equations. We denote the time evolution unitary operator on the interval $\left[0,T\right]$ by $\hat{U}\left(t;s\right)$, which is a solution of 
of the time-dependent Schr{\"o}dinger equation, 
\begin{equation}
i\frac{d}{dt}\hat{U}\left(t;s\right)=\hat{H}\left(t\right)\hat{U}\left(t;s\right),\label{eq:app_time_dep_schro} 
\end{equation}
with $\hat{U}\left(s;s\right)=I$.  We assume that the Hamiltonian has the form $H\left(t\right)=a\left(t\right)\hat{A}+b\left(t\right)\hat{B}$, where the operators $\hat{A}$ and $\hat{B}$ do not commute and that the functions $a\left(t\right)$ and $b\left(t\right)$
are continuous on the interval $\left[0,T\right]$. The Hamiltonian for quantum annealing equation is a special case of this formulation.

Our approach for approximating the solution of Eq.~(\ref{eq:app_time_dep_schro}) as described in Sec.~\ref{sec:background} of the main text
is to combine $N_{M}$ steps of a first order Magnus expansion with
$N_{T}$ steps of the Suzuki-Trotter approximation, i.e.,
\begin{equation*}
\hat{U}\left(T;0\right) \approx \hat{V}\left(T;0\right) = \prod_{k=0}^{N_{M}-1}\hat{V}_{\mathrm{TS}}\left(\left(k+1\right)\Delta t;k\Delta t\right) ,
\end{equation*}
where $\Delta t=\frac{T}{N_{M}}$ and
\begin{align*}
\hat{V}_{\mathrm{TS}}\left(\left(k+1\right)\Delta t;k\Delta t\right) & = \\
& \hspace{-3cm} \left(\exp\left(\frac{-i}{2N_{T}}\intop_{k\Delta t}^{\left(k+1\right)\Delta t}a\left(s\right)ds \hat{A}\right) \right.\\
 & \hspace{-3cm} \times\exp\left(\frac{-i}{N_{T}}\intop_{k\Delta t}^{\left(k+1\right)\Delta t}b\left(s\right)ds \hat{B}\right)\\
 & \hspace{-3cm} \times\left.\exp\left(\frac{-i}{2N_{T}}\intop_{k\Delta t}^{\left(k+1\right)\Delta t}a\left(s\right)ds \hat{A}\right)\right)^{N_{T}} . 
\end{align*}
Let $\hat{V}_{k}$ be the first order Magnus expansion part,
\[
\hat{V}_{k}=\exp\left(-i\intop_{k\Delta t}^{\left(k+1\right)\Delta t} \left( a\left(s\right) \hat{A} + b\left(s\right) \hat{B} \right) ds\right) .
\]
Since $a\left(t\right)$ and $b\left(t\right)$ are bounded on $\left[0,T\right]$,
we have $N_{T}^{-1}\intop_{k\Delta t}^{\left(k+1\right)\Delta t}a\left(s\right)ds=\mathscr{O}\left(\Delta tN_{T}^{-1}\right)$
and $N_{T}^{-1}\intop_{k\Delta t}^{\left(k+1\right)\Delta t}b\left(s\right)ds=\mathscr{O}\left(\Delta tN_{T}^{-1}\right)$.
This implies that the error on the Suzuki-Trotter approximation is \cite[Eq.(1)]{Childs_2021}
\begin{align*}
\left\Vert \hat{V}_{\mathrm{TS}}\left(\left(k+1\right)\Delta t;k\Delta t\right)-\hat{V}_{k}\right\Vert  & =\mathscr{O}\left(\Delta t^{3}N_{T}^{-2}\right) .
\end{align*}
Moreover, the error made by the first order
truncation of the Magnus expansion for a time evolution from $k\Delta t$
to $\left(k+1\right)\Delta t$ is \cite[page 29]{Blanes_2009}
\begin{align*}
\left\Vert \hat{U}\left(\left(k+1\right)\Delta t;k\Delta t\right)-\hat{V}_{k}\right\Vert  & =\mathscr{O}\left(\Delta t^{2}\right).
\end{align*}
This implies that the combined error from Trotter-Suzuki
and first order Magnus during a time step $\Delta t$ is 
\begin{align*}
\left\Vert \hat{U}\left(\left(k+1\right)\Delta t;k\Delta t\right)-\hat{V}\left(\left(k+1\right)\Delta t;k\Delta t\right)\right\Vert \\
=\mathscr{O}\left(\Delta t^{2}\right)+\mathscr{O}\left(\Delta t^{3}N_{T}^{-2}\right)=\mathscr{O}\left(\Delta t^{2}\right)
\end{align*}
Thus, the first order Magnus error is dominating the total approximation error.

Since there are $N_{M}=\frac{T}{\Delta t}$ Magnus steps, the final
error is of order
\begin{equation*}
\left\Vert \hat{U}\left(T;0\right)-\hat{V}\left(T;0\right)\right\Vert =\mathscr{O}\left(N_{M}^{-1}\right) .
\end{equation*}
We see from the above equation that the number of steps for the Suzuki-Trotter formula is not dominating the error and can be chosen as $N_{T}=2$ when $N_{M}$ is large.

\begin{table*}[ht]
\centering
\begin{tabular}{|c|c|c|c|c|c|c|}
\hline
Qubit & T1 $(\mu s)$   & T2 $(\mu s)$   & Frequency (GHz) & Readout Error & P(Meas=0|Prep=1) & P(Meas=1|Prep=0) \\ \hline
0        & 122.754 & 237.973 & 5.071        & 0.023     & 0.014          & 0.009          \\ \hline
1        & 204.498 & 141.381 & 4.930        & 0.015    & 0.009         & 0.006          \\ \hline
2        & 212.978 & 335.274 & 4.670        & 0.011      & 0.006          & 0.004          \\ \hline
4        & 115.938 & 58.890  & 5.021        & 0.019      & 0.012          & 0.007          \\ \hline
\end{tabular}
\caption{Features of the qubits used in the device \texttt{ibmq\_mumbai}  (1.10.0).}
\label{tab:qubit_data_mumbai}
\end{table*}

\begin{table*}
\centering
\begin{tabular}{|c|c|c|c|c|c|c|}
\hline
Qubit  & T1 $(\mu s)$   & T2 $(\mu s)$   & Frequency (GHz) & Readout Error & P(Meas=0|Prep=1) & P(Meas=1|Prep=0) \\ \hline
27                & 422.741                 & 232.809                 & 4.823                   & 0.0274                 & 0.032                    & 0.023                    \\ \hline
28                & 161.740                 & 168.768                 & 4.773                   & 0.0094                 & 0.011                    & 0.008                    \\ \hline
29                & 312.580                 & 234.700                 & 4.699                   & 0.0150                 & 0.017                    & 0.013                    \\ \hline
35                & 278.541                 & 202.674                 & 5.102                   & 0.0256                 & 0.029                    & 0.022                    \\ \hline
\end{tabular}
\caption{Features of the qubits used in the device \texttt{ibm\_osaka}  (1.1.8)  .}
\label{tab:qubit_data_osaka}
\end{table*}

\begin{table}[h!]
\centering
\label{tab:single_osaka}
\begin{tabular}{|c|c|c|c|}
 \hline
\backslashbox{Qubit}{Gate} & $x$ & $sx$ & $id$ \\
\hline
0 & 0.00028 & 0.00028 & 0.00028 \\ \hline
1 & 0.00016 & 0.00016 & 0.00016 \\ \hline
2 & 0.00012 & 0.00012 & 0.00012 \\ \hline
4 & 0.00026 & 0.00026 & 0.00026 \\ \hline
\end{tabular}
\caption{Error gate for single qubit gates of \texttt{ibmq\_mumbai} (1.10.0).}
\end{table}

\begin{table}[h!]
\centering
\label{tab:single_osaka}
\begin{tabular}{|c|c|c|c|}
 \hline
\backslashbox{Qubit}{Gate} & $x$ & $sx$ & $id$ \\
\hline
27 & 0.00014 & 0.00014 & 0.00014 \\ \hline
28 & 0.00053 & 0.00053 & 0.00053 \\ \hline
29 & 0.00011 & 0.00011 & 0.00011 \\ \hline
35 & 0.00177 & 0.00177 & 0.00177 \\ \hline
\end{tabular}
\caption{Error gate for single qubit gates of \texttt{ibm\_osaka}  (1.1.8)  .}
\end{table}

\begin{table}[h!]
\centering
\label{tab:example2}
\begin{tabular}{|c|c|c|c|}\hline
Qubits & Error CX gate & zz (GHz) & jq (GHz) \\ \hline
0\_1 & 0.00694 & -5.58 e-05 & 0.00193 \\ \hline
1\_2 & 0.00958 & -11.9 e-05 & 0.00184 \\ \hline
1\_4 & 0.00528 & -5.11 e-05 & 0.00184 \\ \hline
\end{tabular}
\caption{Two qubit gate error and couplings of \texttt{ibmq\_mumbai} (1.10.0).}
\end{table}

\begin{table}[h!]
\centering
\label{tab:example2}
\begin{tabular}{|c|c|c|c|}\hline
Qubits & Error ECR gate & zz (GHz) & jq (GHz) \\ \hline
27\_28 & 0.00550 & -7.94 e-05 & 0.00200 \\ \hline
28\_29 & 0.00677 & -5.11 e-05 & 0.00197 \\ \hline
28\_35 & 0.00943 & -7.19 e-05 & 0.00199 \\ \hline
\end{tabular}
\caption{Two qubit gate error and couplings of \texttt{ibm\_osaka}  (1.1.8)  .}
\end{table}

\section{Details of the IBM QC devices \texttt{ibmq\_mumbai} and \texttt{ibm\_osaka}.}
\label{apx:lima}

\begin{figure}[t!]
    \centering
    \includegraphics[width=0.485\textwidth]{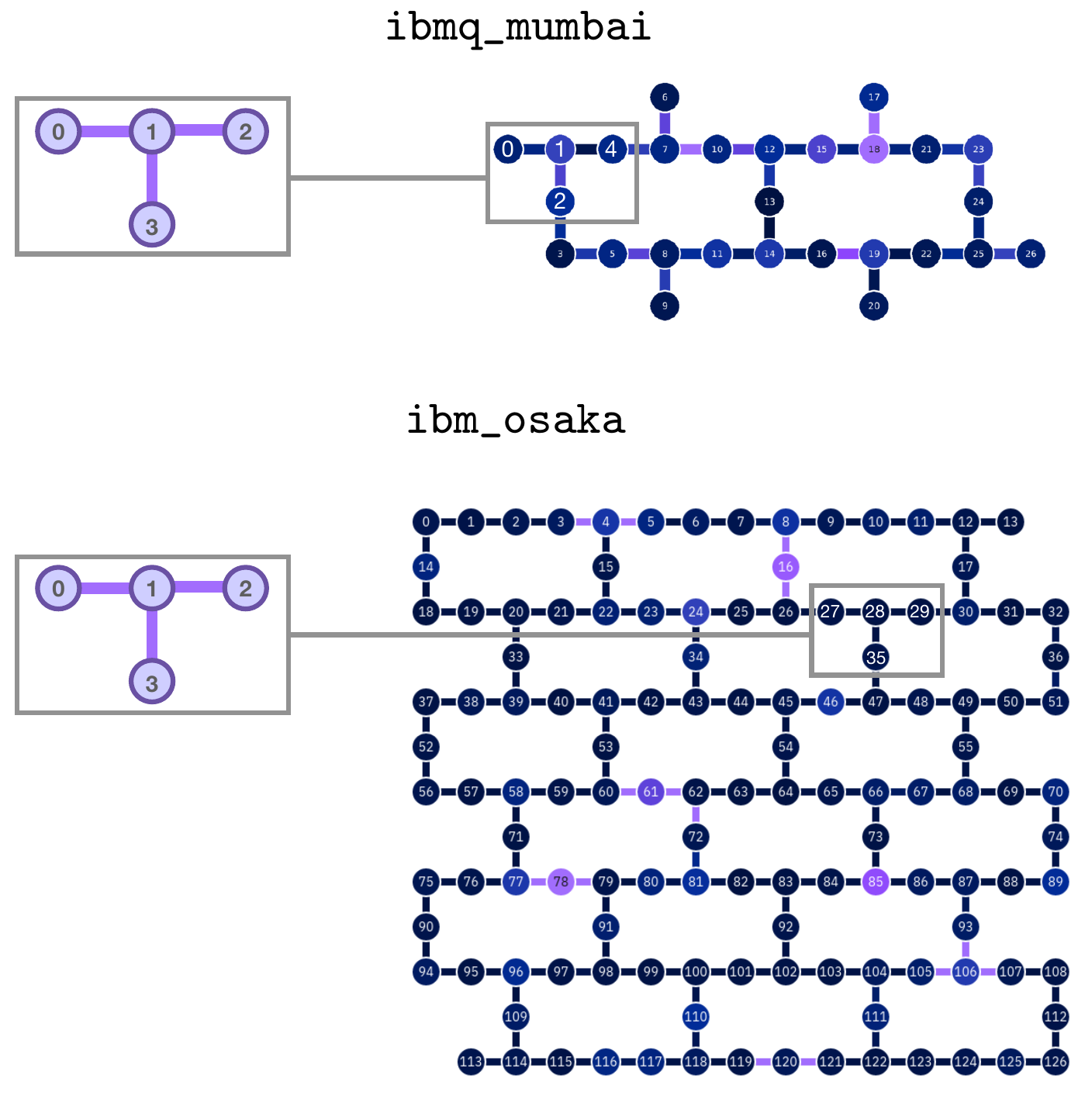} 
    \caption{Embedding of the theoretical model presented in Fig. \ref{fig:HT4-image} into the hardware topology of \texttt{ibmq\_mumbai} and \texttt{ibm\_osaka}.  } \label{fig:Device_embedding}
\end{figure}

In this work, we used the real hardware devices \texttt{ibmq\_mumbai} (1.10.0) and \texttt{ibm\_osaka} (1.1.8) \cite{config_files}, along with the corresponding mock back end for \texttt{ibmq\_mumbai} and an open system simulator incorporating only dephasing in the computational basis and thermal relaxation. The embedding of the theoretical model presented in Fig. \ref{fig:HT4-image} into the hardware topology is presented in Fig. \ref{fig:Device_embedding}

The calibration values of the real hardware devices used for the demonstrations can be found in the Github repository \cite{config_files}. For the simulator, we used the calibration values from \texttt{ibmq\_mumbai} version 1.4.5 \cite{mumbai_qiskit_mock_local}, which were also employed for the mock back end. 

In terms of simulating the physical hardware, the mock back ends use the method \textit{.from\_backend()} \cite{fake_noise} to construct a noise model from calibration data of the physical devices. The model includes: readout error, depolarizing error, and thermal relaxation errors.






\section{Noise calibration for quantum annealing}
\label{sec:analog-open-system}
We model our open analog quantum systems using three different approaches.  The first two are based on using master equations \cite{Lin1976, REDFIELD19651, ame}. 
The first master equation is based on the singular coupling limit (SCL), given by:
\begin{equation}
\frac{d}{dt} \rho(t) = - i \left[ H(t), \rho(t) \right] + \gamma \sum_{i=1}^n \left( \sigma_i^z \rho(t) \sigma_i^z - \rho(t)\right).
\end{equation}
This amounts to having Lindblad operators proportional to the Pauli-$z$ operators, and it describes a pure dephasing model with the dephasing happening in the computational basis.  The characteristic time of this dephasing is given by $1/(2 \gamma)$, which we have set to be $100$ ns.  This corresponds to a choice of $\gamma = 5 \times 10^{-3}$ rad/ns.  To simulate the SCL master equation, we use the HOQST \cite{Chen_2022} implementation of the Lindblad master equation.

For the Adiabatic Master Equation (AME) \cite{ame}, we simulate the open analog quantum dynamics using the HOQST \cite{Chen_2022} implementation.  The AME is a master equation in Lindblad form \cite{Lin1976}, given by:
\begin{eqnarray}
     \frac{1}{\tau}\frac{d}{ds} \rho(s) &=& -i\left[H(s),\rho(s)\right]\nonumber \\
    && + \sum_i \sum_{\omega} \gamma(\omega) \left[ L_{\omega i}(s)\rho(s)L_{\omega i}^{\dagger}(s) \right. \nonumber \\
    && \left.  - \frac{1}{2} \left\{ L_{\omega i}^{\dagger}(s)L_{\omega i}(s), \rho(s) \right\} \right] ,
\end{eqnarray}
where $\gamma(\omega)$ satisfies the KMS condition $\gamma(-\omega) = e^{-\beta \omega} \gamma(\omega)$ \cite{Kub1957,Mar1959,Haa1967}.  The sum over $\omega$ is the sum over all Bohr frequencies, the differences of all possible energy eigenvalues of $H(s)$.  The sum over $i$ is over all system-bath interaction terms. We assume that each qubit interacts with an independent and identical Ohmic heat bath of harmonic oscillators, such that
\begin{equation}
    \gamma(\omega) = 2\pi g^2 \eta \frac{\omega e^{-|\omega|/\omega_c}}{1 - e^{-\beta \omega}},
\end{equation}
where $\omega_c$ is the cutoff frequency, $g^2$ is the system-bath coupling strength, and $\eta$ is a positive constant with dimensions of time-squared arising from the Ohmic bath spectral function. The Lindblad operators are given by
\begin{equation}\footnotesize
    L_{\omega i} = \sum_{a,b}  \delta_{\omega,E_b(s) - E_a(s)} \langle E_a(s)| \sigma_i^z | E_b(s) \rangle |E_a(s)\rangle \langle E_b(s)|,
\end{equation}
where we have assumed a system-bath interaction proportional to $\sigma^z$  for each qubit and ${|E_{a}(s)\rangle}$ are the instantaneous eigenstates of the Hamiltonian with eigenvalues $E_a(s)$.

\begin{figure}[b!]
    \centering
    \includegraphics[width=0.485\textwidth]{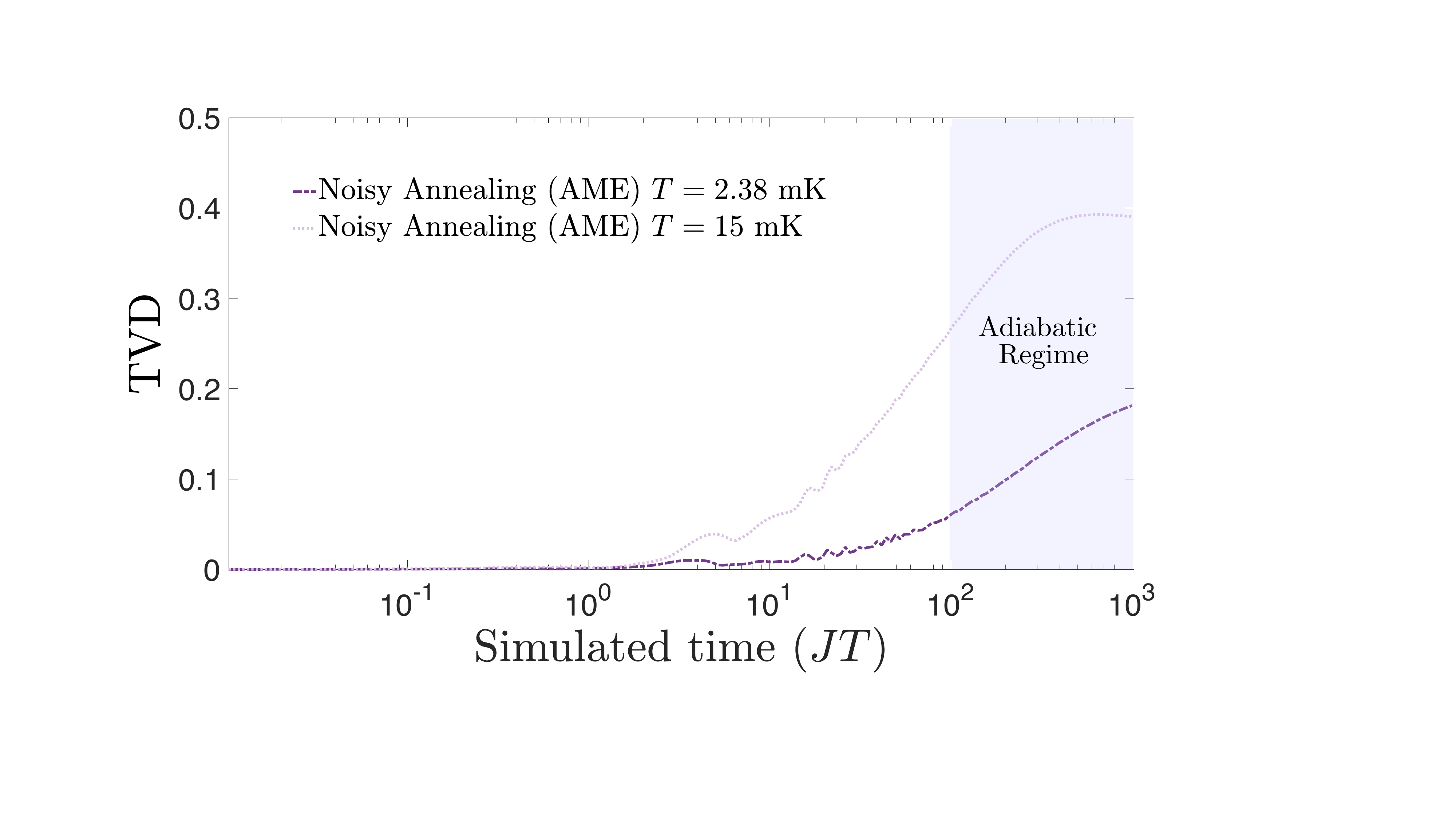} 
    \caption{Comparison of the AME performance at two different temperatures, $T = 15$ and $2.38$ mK, while keeping the dephasing time (Eq.~\eqref{eqt:AMEdephasingtime}) fixed as described in the text.} \label{fig:AME2}
\end{figure}

\begin{figure}[t!]
    \centering
    \includegraphics[width=0.485\textwidth]{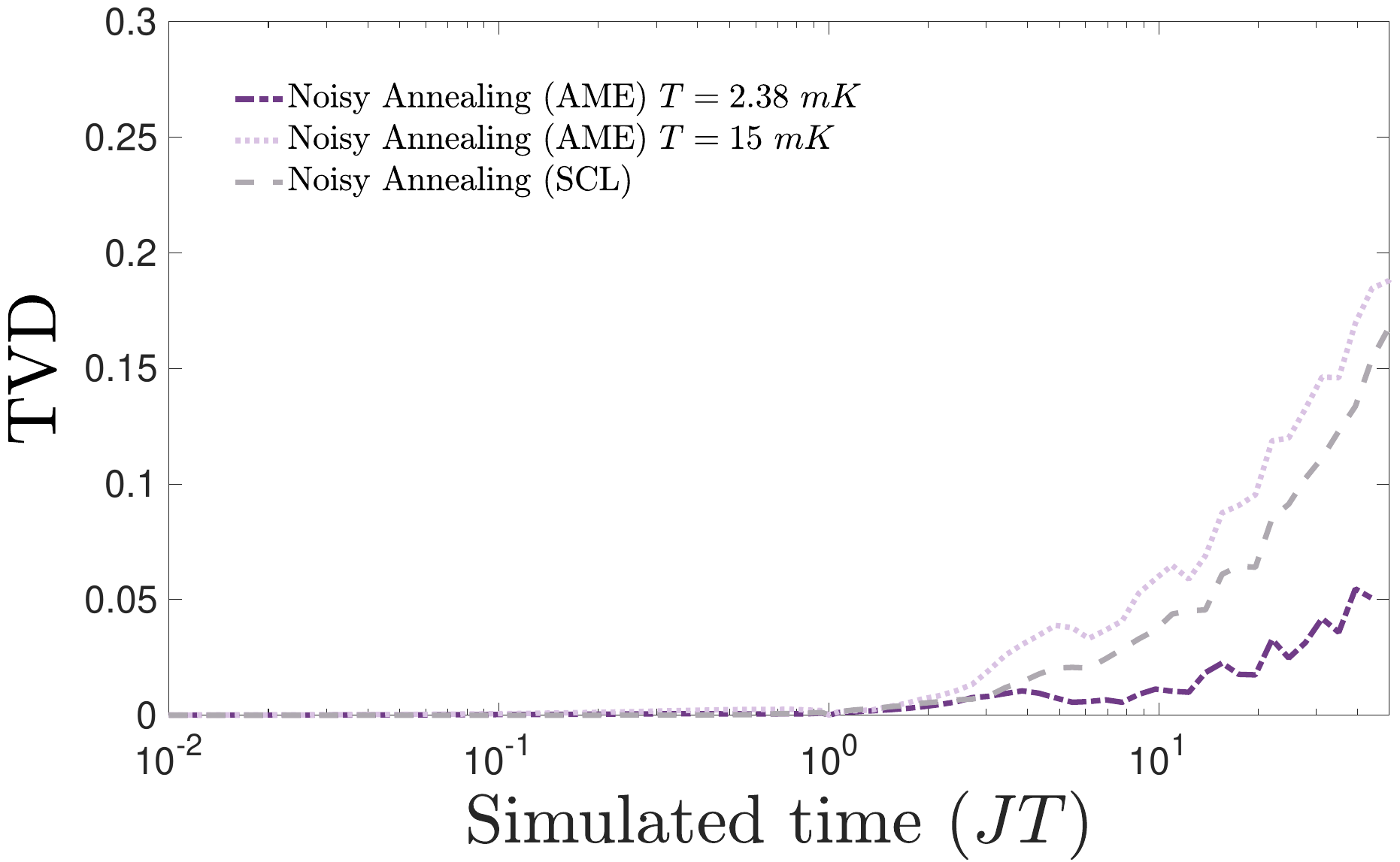}
    \caption{Comparison of the adiabatic master equation (AME) performance at two different temperatures, $T = 15$ and $2.38$ mK and of the singular coupling limit master equation (SCL) including implementation errors, while keeping the dephasing time (Eq.~\eqref{eqt:AMEdephasingtime}) fixed as described in the text. The implementation errors are added as independent Gaussian noise with mean 0 and standard deviation $\sigma = 0.03$ rad/ns to the Ising parameters.} \label{fig:AME3}
\end{figure}

\begin{figure*}
   \includegraphics[width=1\textwidth]{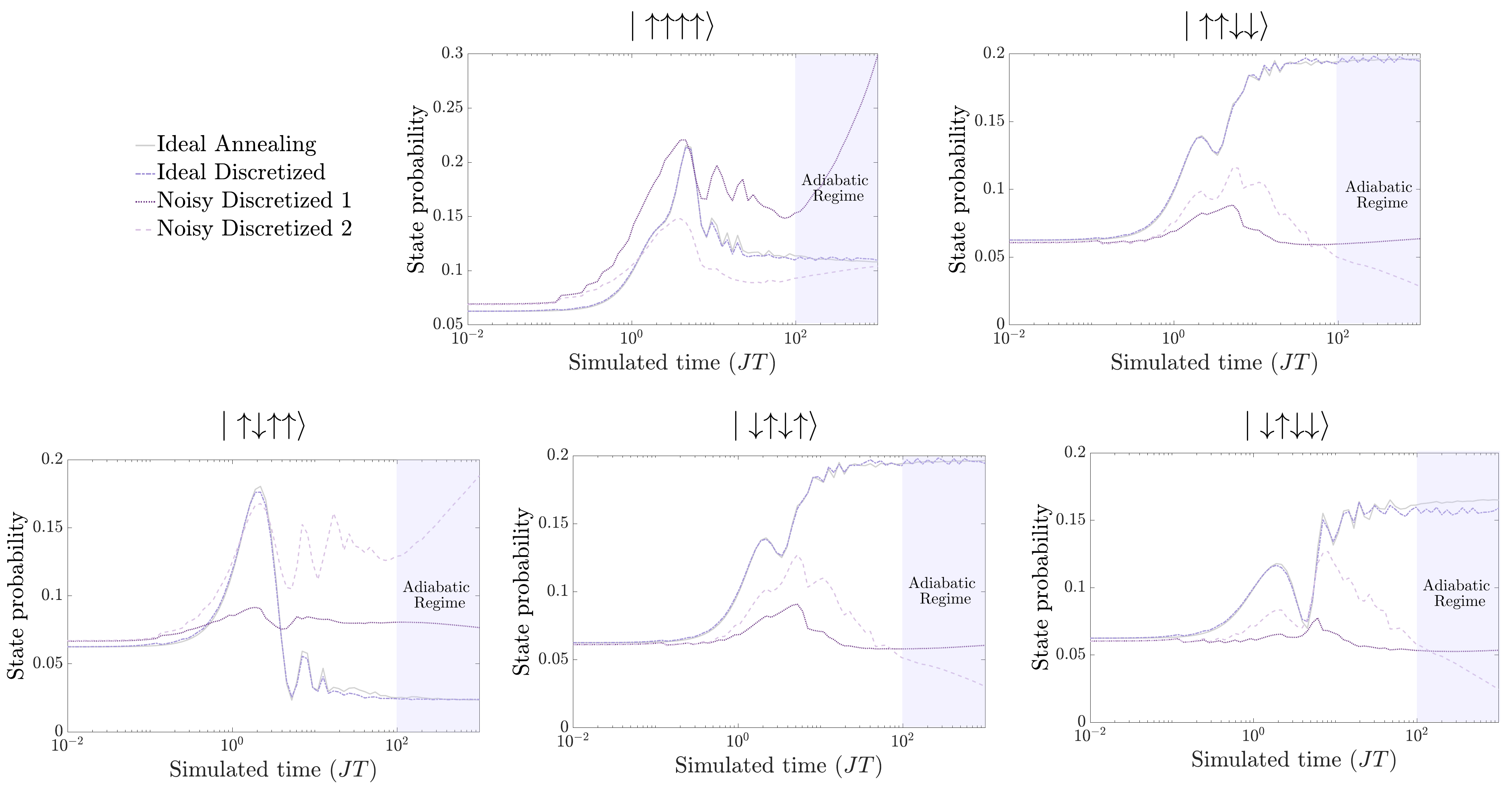} 
    \caption{An analysis of the impact of open system effects on the simulation quality as a function of simulation time $J T$. Ideal Discretized corresponds to simulations with  $N_T = 2$ and the minimum number of Magnus steps $N_M$ needed to achieve a TVD $< 0.01$ in the ideal discrete case. Noisy discretized 1 are open system simulations with only phase damping as described in the main text. Noisy discretized 2 are open system simulations with readout error, depolarizing error and thermal relaxation error as described in the main text. We represent the probability of the different ground state signatures (but $\ket{\uparrow \uparrow \downarrow \uparrow }$) from $\hat{H}_{\mathrm{T}4}$.
}\label{fig:7_appendix}
\end{figure*}
For the purposes of our simulations we use $\omega_c = 8\pi$ rad/ns and a temperature of $15$ and $2.38$ mK corresponding to a temperature energy scale of $1.9643$ and $0.31159$ rad/ns respectively. 
In order to make a fair point of comparison with the SCL, we choose a system-bath coupling ($\eta g^2$) such that the single qubit dephasing time at $s = 1$ is $100$ ns. Using that the single qubit dephasing time is given by \cite{Albash2015}:
\begin{equation} \label{eqt:AMEdephasingtime}
T = \frac{2}{\gamma(\Delta E)(1 + e^{-\beta \Delta E})},
\end{equation}
and using an energy gap of $\Delta E= 2$rad/ns for the linear annealing schedule at $s=1$, we find that $\eta g^2/\hbar^2 = 8.0866 \times 10^{-4}$ for $T = 15$ mK and $\eta g^2/\hbar^2 = 1.7178 \times 10^{-3}$ for $T = 2.38$ mK.

In Fig.~\ref{fig:AME2}, we compare the performance of the AME at the two different temperatures while holding the dephasing time fixed.  We observe  better performance at the colder temperature  at longer times.  This is primarily because the instantaneous thermal state has more weight on the instantaneous ground state, so thermal excitation processes are more suppressed.

Our final open system model is one where we assume coherent evolution for the dynamics but where the implementation of the Ising Hamiltonian is subject to shot-to-shot implementation/programming errors.  To model this kind of noise, we add i.i.d. Gaussian random variables with mean 0 and standard deviation $\sigma = 0.03$rad/ns to the Ising Hamiltonian parameters and simulate the coherent dynamics. We repeat this for $10^3$ noise realizations and average the measured computational basis populations. The behavior of the TVD in this case is shown in Fig.~\ref{fig:Open2} of the main text.  For completeness, we show in Fig.~\ref{fig:AME3} the behavior of the TVD of the adiabatic and singular coupling limit master equations with implementation errors. Comparing to Fig.~\ref{fig:Open2} and ~\ref{fig:AME2}, the addition of implementation errors to our open system master equation simulations does not qualitatively change our results for the choice of noise strengths we have made, indicating that the role of implementation errors in this time range is subdominant.

\section{Additional demonstrations} \label{app:additional}
To supplement the results shown in Fig.~\ref{fig:StateProbabilityOpen} of the main text, we give the probabilities of the remaining ground states in Fig. \ref{fig:7_appendix}.

\end{document}